\documentclass[]{interact}

\usepackage{epstopdf}
\usepackage[caption=false]{subfig}

\usepackage[numbers,sort&compress]{natbib}
\bibpunct[, ]{[}{]}{,}{n}{,}{,}
\renewcommand\bibfont{\fontsize{10}{12}\selectfont}

\theoremstyle{plain}

\theoremstyle{definition}

\theoremstyle{remark}

\begin{document}

\articletype{ORIGINAL RESEARCH PAPER}

\linespread{1.25}

\title{Development and analysis of a Bayesian water balance model for large lake systems}

\author{
\name{Joeseph.~P. Smith\textsuperscript{a}\thanks{CONTACT J.~P. Smith. Tel.: +1-734-741-2252, Fax: +1 734-741-2055, Email: joeseph@umich.edu} and Andrew.~D. Gronewold\textsuperscript{b,c}}
\affil{\textsuperscript{a}Cooperative Institute for Great Lakes Research (CIGLR),
University of Michigan, Ann Arbor, Michigan USA, 48109; ORCID: 0000-0002-1896-1390 \textsuperscript{b}Great Lakes Environmental Research Laboratory, National
Oceanic and Atmospheric Administration, Ann Arbor, Michigan, USA, 48108; \textsuperscript{c}Department of Civil and Environmental Engineering, University
of Michigan, Ann Arbor, Michigan USA, 48109;}
}

\maketitle

\begin{abstract} 

Water balance models (WBMs) are often employed to understand regional hydrologic cycles over various time scales.  Most 
WBMs, however, are physically-based, and few employ state-of-the-art statistical methods to reconcile independent input 
measurement uncertainty and bias. Further, few WBMs exist for large lakes, and most large lake WBMs perform additive 
accounting, with minimal consideration towards input data uncertainty. Here, we introduce a framework for improving a 
previously developed large lake statistical water balance model (L2SWBM). Focusing on the water balances of Lakes Superior
and Michigan-Huron, we demonstrate our new analytical framework, identifying L2SWBMs from 26 alternatives that adequately
close the water balance of the lakes with satisfactory computation times compared with the prototype model. We expect our 
new framework will be used to develop water balance models for other lakes around the world.

\end{abstract}

\begin{keywords}
Bayesian; Network; Markov chain Monte Carlo; Graphical Model; Machine Learning; JAGS; R; Hydrology; Water Balance Model;
Statistical; Analysis; Design of Experiments; Experiment;
\end{keywords}

\section{Introduction}

\begin{figure}
\centering
\includegraphics[width=0.8\textwidth]{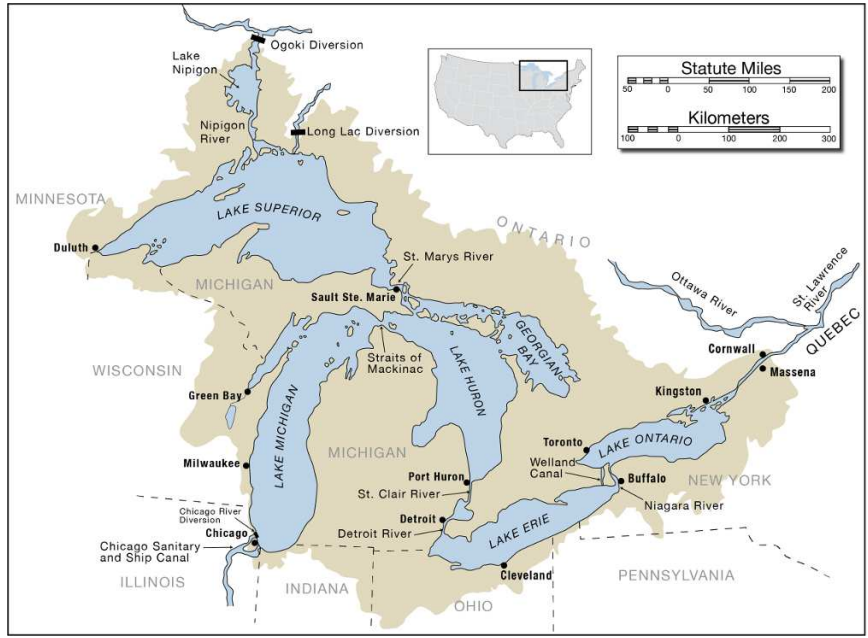}
\caption{The Laurentian Great Lakes and their basin (light brown region) including location of major cities, interbasin 
diversions, and connecting channels.}
\label{fig:gl}
\end{figure}

As global freshwater demand increases \citep{vorosmarty2000global, urbanizationNewFrontier}, there is a growing 
need for a comprehensive understanding of changes and drivers of hydrologic cycles over a wide variety of hydrologic
systems \citep{makhlouf1994two, vorosmarty1998potential, guo2002macro, boughton2004australian, kebede2006water}.  
Water balance models (WBMs) are often employed to understand hydrological systems in numerous practical 
applications including water resources management decision support, and guidance on policies for consumptive use 
and irrigation practices \citep{xu1998reviewOfWaterBalanceMods}. Within a WBM, the balance may be expressed as the flow through a primary river 
-- otherwise known as a rainfall-runoff model \citep{arnell1999simple, mouelhi2006stepwise}, the change in water elevation
of a lake \citep{li2007lake, jin2000application}, or the soil moisture content within an area of focus 
\citep{raes2006simulation, crow2008monitoring}. Modeling at the scale of the large Laurentian 
Great Lakes (figure \ref{fig:gl}), of which Superior and Michigan-Huron are the focus in this publication, is challenging as the lakes 
under study may span multiple countries and data are sparse \citep{swenson2009monitoring}, both in terms of temporal density and spatial coverage.

Existing large lake WBMs typically perform additive accounting with minimal accounting of uncertainty. 
Some water balance components are described by a single source of data while others are described by a single,
potentially biased method which generates estimates. Gibson et al. (2006) \citep{gibson2006hydroclimatic} 
developed an additive WBM of the Great Slave Lake (GSL) in Canada. The GSL WBM was calibrated on observed water levels 
and incoming streamflow estimates which they believed contained the greatest amount of uncertainty. Additionally, the 
GSL WBM incorporated a large measurement record of the outflows through the Mackenzie River, Thiessen weighted 
precipitation gauge estimates, and simulated evaporation. Additive WBMs have 
also been developed for Lake Tana and Lake Victoria. Utilizing 1) precipitation data from Bahr Dar Station near southern Lake Tana, 
2) the Penman equation \citep{penman1948natural} for evaporation given net shortwave radiation data from Addis Ababa Geophysical 
Observatory, and 3) available river discharge data, Kebede et al. (2006) solved an additive water balance equation for Lake Tana 
using Microsoft Excel's (Redmond, Washington, USA) Solver utility \citep{kebede2006water}. Piper et al. (1986) \citep{piper1986water} 
conducted a similar analysis of Lake Victoria, averaging 8 precipitation gauges around the lake, partially simulated 
inflows, and Penman evaporation estimates.

As an enhancement of the additive WBM, we developed a prototype large lake statistical water balance model
(L2SWBM) \citep{gronewold2016HydroDrivers}, seeking an understanding of drivers in the 2013-2014 record-setting water level 
rise on Lakes Superior and Michigan-Huron \citep{GLWLSurgeEOS}, the two largest lakes on Earth by surface area \citep{gron:fort:etal:2013}.
We leveraged state-of-the-art software for Markov chain Monte Carlo (MCMC) simulation of Bayesian Networks to assimilate regional 
estimates of water balance components (see section \ref{sec:data}), developing a first-of-its-kind model that incorporates measurement 
uncertainty, correlation, and bias \citep{kim1996influence, rodell2004basin,  benke2008parameter, sheffield2009closing}, while closing
the water balances of the lakes. We recognize 
that Bayesian WBMs have been developed previously for rainfall-runoff models \citep{bates2001markov}, including WBMs utilizing 
frameworks such as Generalized Likelihood Uncertainty Estimation (GLUE) and the Water And Snow Balance Modeling (WASMOD) 
system \citep{engeland2002bayesian, engeland2005assessing, bulygina2009estimating, jin2010parameter}. 
Prior to our prototype, however, we did not know of other studies that developed a Bayesian WBM for large lakes incorporating multiple independent 
estimates of water balance components. 

Additionally, in developing a prototype L2SWBM, we knew of limited, if any, methodological guidance for developing and selecting a suitable 
WBM in light of criteria relevant to water resource management agencies, policy makers, and other 
model end-users. A lack of guidance in L2SWBM development may have enabled development of a computationally expensive prototype.
Our prototype Superior and Michigan-Huron L2SWBM described the lakes' water balances as cumulative sums of monthly changes in water level 
or height ($\Delta{H_{l,t,C}}$), a proxy for water storage, starting at a specified base month. For the prototype, the base month was January of 
2005, with the period of study ending December 2014, for a total of T = 120 months. At each time step, $t$, the prototype model aggregated 
all components of the water balance of each lake, $l$, from the base month through $t$, conceptually following the 
continuous time, or long-term yield, model of the Soil and Water Assessment Tool (SWAT \cite{arnold1998large, joseph2013using}):

\begin{equation}
\label{eq:pwb}
    \Delta{H_{l,t,C}}  =  \sum_{i=1}^{t}(P_{l,i} - E_{l,i} + R_{l,i} + I_{l,i} - Q_{l,i} \pm D_{l,i} + \epsilon_{l,i}), t \in [1,T]
\end{equation}

\noindent where (all in mm over each lake surface area) $\Delta{H}$ is assumed to be
the difference between water levels at the beginning of months $1$ and $t+1$, $P$ is over-lake precipitation, $E$ is over-lake
evaporation, $R$ is lateral tributary runoff into the lake, $I$ is inflow from an upstream connecting
channel, $Q$ is outflow to a downstream connecting channel, $D$ represents diversions into or
(expressed as a negative value) out of the lake basin, and $\epsilon$ is a process error \citep{ahrestani2013importance} term accounting
for thermal expansion, glacial isostatic rebound, groundwater fluxes, and other sources of variability
in monthly water levels not explained by water balance components $P, E, R, I, Q$, and $D$ alone, nor are
consistently measured in terms of impact on water level. The prototype L2SWBM, according to results in this manuscript,
can take roughly 16 hours to simulate the network and acquire reliable water balance component estimates.

While the prototype L2SWBM differentiated hydrologic drivers of the 2013-2014 water level rise on Lakes Superior and Michigan-Huron 
under a closed water balance \citep{constrainedKalman_pan_wood_2006} for the 120 months of analysis, regional water resource management 
authorities have expressed interest in an expanded 
version of the model applied to all of the Laurentian Great Lakes and across a longer historical period -- about 65 years 
(or 780 months), compared to just 10 (or 120 months) for the prototype. Long run times hinder exploration and formal assessment 
of different model formulations, and a 16 hour run time is problematic considering that the prototype model only includes 
Lake Superior and Lake Michigan-Huron for the 120 months between 2005 to 2014. Our need to explore different model formulations
arises, in part, from water managers' comments that St.\ Clair River flows inferred from the prototype L2SWBM
are biased with a broad, unrealistic range of uncertainty, and that future L2SWBM development should reflect 
stronger \emph{a priori} opinions about the accuracy of \emph{in situ} measurements in the St.\ Clair and other 
connecting rivers, or connecting channels \citep{quin:guer:1986, muel:abad:garc:etal:2007}. Prototype L2SWBM channel flow 
inference bias may reflect unresolved uncertainties in individual lake water balance component estimates, which may propagate
and accumulate through the Laurentian Great Lakes hydrologic system as represented by Bayesian networks. In essence, our goal is
to limit the amount of cumulative uncertainty from Lake Superior through Lake Ontario and the St. Lawrence River in an
expanded version of the L2SWBM.

The objective of this study, therefore, is the development of a framework for systematic experimentation,
evaluation, and selection of alternative formulations of an L2SWBM.  
In this manuscript, we applied the framework to the prototype L2SWBM for Lakes Superior and Michigan-Huron to demonstrate how it 
can be used to improve model efficiency while incorporating \emph{a priori} opinions about biases and uncertainties in 
existing data sources for components of the Great Lakes hydrologic cycle. We expect the new framework will
be useful to improve not just the Great Lakes L2SWBM, but also variations for other large lakes of the world, and that the resulting
new Great Lakes L2SWBM, following implementation of our recommended improvements, will be
suitable for deployment in operational environments and for expansion across the entire Great
Lakes system over a longer historical period.

\subsection{Data}\label{sec:data}

\begin{table*}
\centering
\resizebox{0.95\textwidth}{!}{
\begin{tabular}{ c  l  c  c  c  c  c  c }
Variable                 & \multicolumn{1}{c}{Data source}                & Year range used & Mean & Median & S.D. & 2.5\% & 97.5\% \\       
\hline \hline                                                                             
$y_{SUP,\Delta{H},t,1}$  & CCGLBHHD \cite{ccghbhhd:1977}                  & 2005-2014 & 0.19 & -10 & 72.07 & -110 & 150 \\ 
$y_{SUP,\Delta{H},t,12}$ & CCGLBHHD \cite{ccghbhhd:1977}                  & 2005-2014 & -1.69 & 0 & 152.88 & -300 & 320 \\ 
$y_{SUP,\Delta{H},t,C}$  & CCGLBHHD \cite{ccghbhhd:1977}                  & 2005-2014 & NA & NA & NA & NA & NA \\ 
\hline
$y_{SUP,P,1,t}$          & GLM-HMD \cite{hunter2015development}          & 1950-2014 & 65.53 & 61.65 & 28.13 & 21.60 & 128.05 \\ 
$y_{SUP,P,2,t}$          & CaPA \cite{lespinas2015performance}           & 2005-2014 & 75.27 & 74.32 & 28.26 & 29.00 & 134.57 \\ 
\hline
$y_{SUP,E,1,t}$          & GLM-HMD (LLTM) \cite{hunter2015development}   & 1950-2014 & 46.06 & 40.69 & 42.18 & -4.96 & 128.79 \\ 
$y_{SUP,E,2,t}$          & GEM-MESH \cite{deacu2012predicting}           & 2005-2014 & 48.85 & 45.56 & 41.18 & -7.74 & 122.08 \\ 
\hline
$y_{SUP,R,1,t}$          & GLM-HMD (ARM) \cite{hunter2015development}    & 1950-2014 & 47.07 & 37.63 & 27.20 & 18.24 & 119.33 \\ 
$y_{SUP,R,2,t}$          & NOAA-GLERL LBRM \cite{croley2005distributed}  & 1950-2014 & 50.77 & 44.15 & 23.41 & 23.24 & 111.27 \\ 
\hline
$y_{SUP,Q,1,t}$          & CCGLBHHD \cite{ccghbhhd:1977}                 & 1950-2014      & 69.71 & 66.44 & 16.67 & 47.65 & 108.38 \\ 
$y_{SUP,Q,2,t}$          & IGS \cite{usgsGauges}                         & Nov. 2008-2014 & 58.20 & 51.61 & 16.31 & 37.60 & 100.50 \\ 
\hline
$y_{SUP,D,1,t}$          & CCGLBHHD \cite{ccghbhhd:1977}                 & 1950-2014 & 4.91 & 4.22 & 2.59 & 1.27 & 11.21 \\ 
\hline\hline
$y_{MHU,\Delta{H},t,1}$  & CCGLBHHD \cite{ccghbhhd:1977}                 & 2005-2014 & 0.80 & -10 & 72.65 & -120 & 155.25 \\ 
$y_{MHU,\Delta{H},t,12}$ & CCGLBHHD \cite{ccghbhhd:1977}                 & 2005-2014 & 1.46 & -20 & 250.49 & -470 & 488 \\ 
$y_{MHU,\Delta{H},t,C}$  & CCGLBHHD \cite{ccghbhhd:1977}                 & 2005-2014 & NA & NA & NA & NA & NA \\ 
\hline
$y_{MHU,P,1,t}$          & GLM-HMD \cite{hunter2015development}          & 1950-2014 & 70.16 & 68.25 & 27.16 & 24.35 & 132.99 \\ 
$y_{MHU,P,2,t}$          & CaPA \cite{lespinas2015performance}           & 2005-2014 & 81.09 & 79.14 & 28.75 & 35.40 & 140.73 \\ 
\hline
$y_{MHU,E,1,t}$          & GLM-HMD (LLTM) \cite{hunter2015development}   & 1950-2014 & 42.68 & 35.12 & 37.36 & -4.53 & 118.31 \\ 
$y_{MHU,E,2,t}$          & GEM-MESH \cite{deacu2012predicting}           & 2005-2014 & 63.18 & 61.53 & 44.47 & 0.93 & 146.12 \\ 
\hline
$y_{MHU,R,1,t}$          & GLM-HMD (ARM) \cite{hunter2015development}    & 1950-2014 & 60.73 & 53.1 & 31.81 & 22.22 & 137.90 \\ 
$y_{MHU,R,2,t}$          & NOAA-GLERL LBRM \cite{croley2005distributed}  & 1950-2014 & 62.26 & 56.40 & 28.66 & 25.32 & 131.77 \\ 
\hline
$y_{MHU,Q,1,t}$          & CCGLBHHD \cite{ccghbhhd:1977}                 & 1950-2014 & 119.77 & 119.42 & 14.26 & 90.39 & 144.42 \\ 
$y_{MHU,Q,2,t}$          & IGS \cite{usgsGauges}                         & Nov. 2008-2014 & 112.19 & 113.74 & 10.11 & 86.13 & 128.42 \\ 
\hline
$y_{MHU,D,1,t}$          & CCGLBHHD \cite{ccghbhhd:1977}                 & 1950-2014 & 2.05 & 2.02 & 0.56 & 1.16 & 3.22 \\       
\hline
\end{tabular}
}\caption{Summary of data sets used to develop water balance component prior probability
distributions and as a basis for likelihood functions. Units for statistics -- mean, median, standard deviation (S.D.), 2.5\% 
and 97.5\% quantiles -- are millimeters over the respective lake surface.
Unless otherwise specified, year range specified for each data set indicates availability from January of the first year 
through December of the second year. See table \ref{tab:symbols} for definitions of variables in the first column.}\label{tab:data}
\end{table*}

We used monthly, 1-dimensional (depth over lake surface), time-series data from a variety of independent 
sources to develop and test the Great Lakes L2SWBM (table \ref{tab:data}).
Data are used specifically in deriving water balance component prior probability distributions and likelihood functions
described in sections \ref{sec:wbf} and \ref{sec:wbct}. Data sources include the 
National Oceanic and Atmospheric Administration's Great Lakes Environmental Research Laboratory (NOAA-GLERL), 
Environment and Climate Change Canada (ECCC), and the Coordinating Committee 
on Great Lakes Basic Hydraulic and Hydrologic Data (CCGLBHHD, hereafter referred to as the Coordinating Committee), 
an ad hoc group of science agencies from both the United States and Canada.

NOAA-GLERL has, for many years, developed the Great Lakes Monthly Hydrometeorological Database (GLM-HMD)
\cite{hunter2015development}. The GLM-HMD utilizes a suite of models to generate 1-dimensional  
estimates of precipitation, evaporation, and runoff. GLM-HMD precipitation estimates are derived from Thiessen weighting of meteorological station 
precipitation estimates across the Great Lakes basin \cite{croley1985resolving}. 1-dimensional estimates of evaporation in the GLM-HMD 
utilize regional meteorological measurements of wind speed, dew point, cloud cover, and temperature input into the Large 
Lake Thermodynamics model (LLTM, \cite{croley1989verifiable, croley1992long, thermoModel}). United States Geological Survey (USGS) and 
Water Survey Canada (WSC) streamflow estimates across the basin are aggregated into GLM-HMD estimates of runoff for each lake via a 
conventional Area-Ratio Method (ARM) \cite{fry2013identifying}. Lastly, in addition to the ARM runoff model estimates, we utilized runoff estimates
from the Large Basin Runoff Model (LBRM)\cite{croley2005distributed}, which simulates water movement in a watershed 
through two soil layers (upper and lower), groundwater deposits, and evaporation, using meteorological station
temperature and precipitation measurements as inputs.

Additional 1-dimensional, over-lake precipitation estimates are derived from gridded outputs of Environment and Climate Change Canada's 
(ECCC) version of the Canadian Precipitation Analysis (CaPA). CaPA utilizes short-term numerical weather prediction (NWP)
models, along with meteorological station precipitation estimates from a collection of networks across Canada and the United 
States. NWP models depend on the Global Environmental Multiscale (GEM) model, which along with the  Mod\'{e}lisation Environmentale 
- Surface et Hydrologie's (MESH) surface model, produce additional evaporation estimates utilized in this paper \cite{lespinas2015performance, 
deacu2012predicting}. 

The Coordinating Committee produces daily surface water elevation, or water level estimates, which we used to compute 
observations of $\Delta{H}$. Water levels for the Laurentian Great Lakes are estimated via a series of water level gauges around the 
coasts maintained by NOAA National Ocean Service's Center for Operational Oceanographic Products and Services (NOAA/NOS CO-OPS) and the 
Canadian Department of Fisheries and Oceans' Canadian Hydrographic Service (DFO-CHS). 
To adjust for geological phenomena such as isostatic rebound, or the continued expansion of the Earth's crust from glacial retreat, 
water level estimates reference the International Great Lakes Datum (IGLD).

Additionally, the Coordinating Committee has historically developed connecting
channel and diversion flow estimates within the Great Lakes basin. Connecting channel flows are estimated using a variety of methods, 
depending on the physical environment at the outlet of each lake. Flow estimates for the St. Marys river connecting Lakes 
Superior and Michigan-Huron, for example, are computed from the flows through a collection of dams and marine navigation locks between
Sault Ste.\ Marie, Michigan in the United States, and Sault Ste.\ Marie, Ontario, Canada. 
For the St. Clair River, a combination of Acoustic Doppler Velocity Meter (ADVM) measurements and stage-fall discharge equation estimates produce 
the time-series data implemented in the L2SWBM. As a second source of channel flow estimates, ADVMs are used within International Gauging Stations 
(IGS). IGS used in this study are located at Sault Ste.\ Marie (for outflows from Lake Superior)
and Port Huron (for outflows from Lake Michigan-Huron) and are maintained through a partnership
between the USGS and WSC. 

Measurements of diversions into (or out of) each lake basin include the Ogoki River and Long-Lac
diversions into Lake Superior, and the Chicago River diversion out of Lake Michigan-Huron.
We assumed that channel flows and diversions make their contribution to the water balance of a lake
at the lake's outlet, downstream lake's inlet, or the point where a diversion enters or exits a lake.
A notable challenge to this assumption includes the Ogoki diversion. Water diverted
from the Ogoki River must first pass through Lake Nipigon (northern-most lake in figure \ref{fig:gl}) 
before it arrives in Lake Superior, thus there is an unknown amount of residence time before the water 
is actualized in the Lake Superior balance. As the diversion is
small compared to other components of the water balance, however, we assume the impact uncertainty has 
on the water balance is minimal.

For further reading on coordinated estimates of water level, channel flows, and diversions, see \citep{gronewold2016HydroDrivers}.

\section{Methodology}\label{sec:methods}

\begin{table*}
\centering
\resizebox{0.95\textwidth}{!}{
\begin{tabular}{ l l }
Symbol                  & Description              \\       
\hline \hline
\multicolumn{2}{l}{Subscripts and related variables} \\
\hline
$t$             & An individual month, spans $[1,T]$ \\
$T$             & Total months in analysis, 120 from January 2005 through December 2014 in this manuscript \\
$c(t)$          & Calendar month $[1,12]$ that month $t$ is in \\
$w$             & Rolling window for water balance analysis \\
$j$             & Start month for an individual water balance window spanning $[1,T-w+1]$\\
$n$             & Independent estimate of $\theta$ \\
$l$             & An individual lake, either SUP (Superior) or MHU (Michigan-Huron) in this manuscript \\
\hline                                                          
\multicolumn{2}{l}{True, but unobserved parameters (mm over lake surface)} \\
\hline
$\Delta{H}_{l,j,w}$  & Change in water level for lake $l$ from beginning of month $j$ to beginning of month $j+w$ \\
$\Delta{H}_{l,t,C}$	 & Change in water level for lake $l$ from beginning of month $1$ to beginning of month $t$  \\
$P_{l,t}$	           & Precipitation (total) over lake $l$ in month $t$ \\
$E_{l,t}$	           & Evaporation (total) from lake $l$ in month $t$ \\
$R_{l,t}$	           & Basin runoff (total) into lake $l$ in month $t$ \\
$I_{l,t}$	           & Connecting channel inflow (total) for lake $l$ in month $t$ \\
$Q_{l,t}$	           & Connecting channel outflow (total) for lake $l$ in month $t$ \\
$D_{l,t}$	           & Diversion from or to a lake (total) for lake $l$ in month $t$ \\
$\theta$             & Used to represent $P, E, R, Q, and \ D$ \\
\hline
\multicolumn{2}{l}{Observed variables for likelihoods (mm over lake surface)} \\
\hline
$y_{l,H,t}$            & Water level estimate for lake $l$ at beginning of month $t$ \\
$y_{l,\Delta{H},j,w}$  & Coordinated estimate of $\Delta{H}_{l,j,w}$ \\
$y_{l,\Delta{H},t,C}$  & Coordinated estimate of $\Delta{H}_{l,j,C}$ \\
$y_{l,\theta,n,t}$     & $n_{th}$ Independent estimate of $\theta_{l,t}$ \\                  
\hline
\multicolumn{2}{l}{Water balance component prior distribution parameters} \\
\hline
$\hat{\mu}_{l,E,c(t)}, \hat{\mu}_{l,Q,c(t)}, \hat{\mu}_{l,D,c(t)}$ & Mean of historical data for E, Q, and D by calendar month $c(t)$ and lake $l$\\
$\hat{\mu}_{l,ln(R),c(t)}$ & Mean of natural logarithm of historical data for R by calendar month $c(t)$ and lake $l$\\
$\hat{\tau}_{l,E,c(t)}, \hat{\tau}_{l,Q,c(t)}, \hat{\tau}_{l,D,c(t)}$ & Precision of historical data for E, Q, and D by calendar month $c(t)$ and lake $l$\\
$\hat{\tau}_{l,ln(R),c(t)}$ & Precision of natural logarithm of historical data for R by calendar month $c(t)$ and lake $l$\\
$\psi_{1,l,c(t)}, \psi_{2,l,c(t)}$ & Shape and rate parameters for the prior distribution for P by calendar month $c(t)$ and lake $l$\\
\hline
\multicolumn{2}{l}{Hyperparameters} \\
\hline
$\tau_{l,\theta,n}$    & Precision of $y_{l,\theta,n,t}$ for all $t$ \\
$\tau_{l,\Delta{H},w}$ & Precision of $y_{l,\Delta{H},j,w}$ for all $j$ \\
$\tau_{l,\Delta{H},C}$ & Precision of $y_{l,\Delta{H},t,C}$ for all $t$ \\
$\epsilon_{l,t}$       & Process error in water balance equation for lake $l$ in month $t$ in mm over lake surface \\
$\epsilon_{l,c(t)}$    & Seasonal process error by calendar month $c(t)$ \\
$\tau_{l,\epsilon,c(t)}$ & Seasonal precision of monthly process error by calendar month $c(t)$ \\
$\eta_{l,\theta,n,t}$    & Bias of $y_{l,\theta,n,t}$ in mm over lake surface \\
$\eta_{l,\theta,n,c(t)}$ & Seasonal bias of $y_{l,\theta,n,t}$ by calendar month $c(t)$ \\
$\tau_{l,\eta,c(t)}$     & Seasonal precision of $y_{l,\theta,n,t}$ bias by calendar month $c(t)$ \\
\hline
$\pi(...)$             & Prior distribution for variable or parameter \\
$\tau$                 & Precision $1/\sigma^{2}$, an expression of variance, $\sigma^{2}$, and standard deviation, $\sigma$ \\
\end{tabular}
}\caption{Summary of variables and parameters employed in this manuscript}\label{tab:symbols}
\end{table*}

Given the available data, we discuss the modeling frameworks, or Bayesian networks, we employed to infer values for each 
lake's water balance in this section. Further, we define a new water balance formulation designed to increase efficiency 
relative to the prototype L2SWBM. Lastly, we review parameter structures of the L2SWBM throughout this section, 
along with combinations of L2SWBM parameter structures that we tested in developing the model. For convenience, definitions of
all variables and parameters in this section are in table \ref{tab:symbols}.

\subsection{Modeling framework}

\subsubsection{Encoding the Bayesian Network}

With all variations of the L2SWBM, we inferred values for each component of each lake's monthly water balance, 
along with the other model parameters (e.g. process error described in section \ref{sec:wbf}), through statistical 
models known as Bayesian networks \cite{malve2007bayesian, dorner2007multi}. We programmed 
the Bayesian networks in the BUGS modeling language (Bayesian inference Using Gibbs Sampling 
\cite{lunn2000winbugs, lunn2009bugs}), describing prior distributions, likelihoods, and posterior predictive 
distributions highlighted in all of section \ref{sec:methods}. For other examples of BUGS applications, see Armero et al. (2008)
\citep{armero2008bayesian} and Blangiardo and Baio (2014) \citep{blangiardo2014evidence}. 
To compile and sample the complete likelihood function, joint, and conditional posterior 
distributions of the Bayesian Networks, we used JAGS \citep[Just Another Gibbs Sampler; see][]{plummer2003jags}, and the 
`rjags' package in the R statistical software environment \citep{team2013r}. JAGS, despite what its name implies,
utilizes a variety of samplers, and for our models, JAGS utilized Gibbs and slice samplers. Slice samplers are used 
on distributions describing non-negative, asymmetrically distributed variables, being precipitation and 
runoff (see section \ref{sec:priors}).

\subsection{Water balance}\label{sec:wbf}

For our new, experimental L2SWBMs, instead of a cumulative water balance (equation \ref{eq:pwb}), we programmed a formulation in which 
changes in lake storage on lake $l$ are considered across a period of $w$ months. $\Delta{H}_{l,j,w}$ is defined 
as the difference between the lakewide-average surface water elevation
at the beginning of month $j$, and the lakewide-average surface water elevation at the beginning
of month $j+w$ (we hereafter refer to $w$ as the length, in months, of the water balance
rolling window). Changes in lake storage are then related to monthly-total water balance
components as follows:

\begin{equation}
\label{eq:wb}
    \Delta{H}_{l,j,w} = H_{l,j+w} - H_{l,j} = \sum_{i=j}^{j+w-1}(P_{l,i} - E_{l,i} + R_{l,i} + I_{l,i} - Q_{l,i} + D_{l,i} + \epsilon_{l,i})
\end{equation}

\noindent We note $I_{SUP,t} = 0$ because Lake Superior is the most upstream of the 
Laurentian Great Lakes and there is no inflow from a connecting channel. $I_{MHU,t} = 0.7(Q_{SUP,t})$, 
where 0.7 is a scaling factor accounting for the difference 
between surface areas of the two lakes. Through $Q_{SUP,t}$ and $I_{MHU,t}$, we linked the lakes in the model
and ensured consistent inferences of flows through the St. Marys river.

Observations $y_{l,\Delta{H},t,w}$ were computed from coordinated beginning of month
estimates, and linked to the true value $\Delta{H}_{l,j,w}$ (equation \ref{eq:wb}) via the likelihood:

\begin{eqnarray}
		y_{l,H,j+w} - y_{l,H,j} = y_{l,\Delta{H},j,w} & \sim & \mathsf{N}(\Delta{H}_{l,j,w}, \tau_{l,\Delta{H},w})\label{eq:wlobs}
\end{eqnarray}

\noindent where precision $\tau_{l,\Delta{H},w}$ is given a vague $\mathsf{Gamma}(0.01,0.01)$ prior. 
No variations of this formulation were considered. 

We recognize identifiability \citep{eberly2000identifiability, rannala2002identifiability, gill2007partial, 
renard2010understanding} is a challenge with the additive balance in equation \ref{eq:wb} being a part of
the L2SWBM. In fact, the challenge of identifiability in WBMs was indirectly highlighted by Piper et al. (1986) \citep{piper1986water},
who noted that ``it is not possible to distinguish easily between underestimates of rainfall and over-estimates of evaporation''. 
By analyzing convergence of our experimental L2SWBMs (elaborated upon in section \ref{sec:eval}), 
we discover identifiable models, even though defining identifiability is difficult in a Bayesian context,
where inferences are best given a model and data, not just a model \citep{gelmanIDBayes}. The odds of finding
an identifiable model are thus increased given strong prior distributions for the components $\theta$, component
estimates' biases $\eta$, and process error $\epsilon$ (described later in this section), along with prior distributions
for estimates' precisions $\tau$ that grant more weight to other parameters of the network. Prior specifications for $\theta$,
$\eta$, and $\tau$ are described in section \ref{sec:wbct}.

In discovering an identifiable L2SWBM, we compared three sets of models in the experiment design detailed 
in section \ref{sec:design}, where each set has a different water balance formulation. The first set
uses the balance computation in equation \ref{eq:pwb} as a basis of comparison. Observations of change
in storage in the first set of models are denoted $y_{l,\Delta{H},t,C}$ in table \ref{tab:data}, where $C$ indicates
a cumulative water balance over the analysis period. Further, $\Delta{H}_{l,t,C} = H_{l,t+1} - H_{l,1}$, $i$ spans $[1,t]$
in the summation in equation \ref{eq:wb}, and $y_{l,H,t+1} - y_{l,H,1} = y_{l,\Delta{H},t,C}$ for equation \ref{eq:wlobs}.
The other two sets of experimental models with respect to water balance formulation use equation \ref{eq:wb} with rolling windows 
of $w = 1$ and $w = 12$ months. We not only intend to 
demonstrate a reduction in model computation time with the latter two sets in contrast to the first, but
also assess the impact of different rolling window lengths on water balance estimates (e.g.\ monthly, 
seasonal, and inter-annual scales) relevant to regional water resource management decisions.

\subsubsection{Process error}

Additionally, we considered three alternative structures for process error $\epsilon_{l,t}$.
The first, and simplest, is where we assume $\epsilon_{l,t} = 0$ as in the prototype model's implementation and unexplained variability in observed 
changes in storage propagates into water level measurement uncertainty (equation \ref{eq:wlobs}) and the uncertainty of 
other water balance components (see section \ref{sec:wbct}). The second structure for process error strictly tracks
potential seasonal variation, mapping each $\epsilon_{l,t}$ to one of twelve `fixed', calendar-month $c(t)$ specific terms such that:

\begin{eqnarray}
	\epsilon_{l,t} & = & \epsilon_{l,c(t)}\notag\\ 
	\pi(\epsilon_{l,c(t)}) & \sim & \mathsf{N}(0,0.01)\label{eq:error1}
\end{eqnarray}

\noindent where variance ($\sigma^{2} = 100$) is expressed in terms of precision ($\tau = 1/\sigma^{2}$ = 0.01)
in the vague, zero-mean prior. The third and final alternative structure for process error is hierarchical, where each
$\epsilon_{l,t}$ is given a vague prior. For each monthly process error prior, the mean is mapped to a calendar-month-specific 
term as in equation \ref{eq:error1}, while the precision is also calendar-month-specific, with all twelve precisions given a 
vague prior:

\begin{eqnarray}
 \pi(\epsilon_{l,t}) 					& \sim  & \mathsf{N}(\epsilon_{l,c(t)}, \tau_{l,\epsilon,c(t)}) \label{eq:error2} \\
 \pi(\epsilon_{l,c(t)})       & \sim  & \mathsf{N}(0,0.01)\notag \\
 \pi(\tau_{l,\epsilon,c(t)})  & \sim  & \mathsf{Gamma}(0.05,0.05)\notag
\end{eqnarray}

\noindent Alternative structures for $\epsilon$ represented in equations \ref{eq:error1} and \ref{eq:error2} are intended to 
isolate water level measurement uncertainty from uncertainty in the balance model. We therefor modeled $\epsilon$ at 
seasonal and monthly temporal resolutions, respectively, while $\tau_{l,\Delta{H},w}$
(equation \ref{eq:wlobs}) is assumed to be constant throughout the analysis period.

\subsection{Water balance component ($\theta$) priors and likelihoods}\label{sec:wbct}

We treat water balance components ($\theta \in [P, E, R, Q, D]$) as random variables in the L2SWBM, attributing
informative prior distributions and likelihoods, the latter of which we link input independent
estimates. In this sub-section we first describe the prior distributions we develop, and then 
the likelihoods.

\subsubsection{Prior distributions}\label{sec:priors}

\begin{figure}
\centering
\includegraphics[angle=270, width=0.9\textwidth]{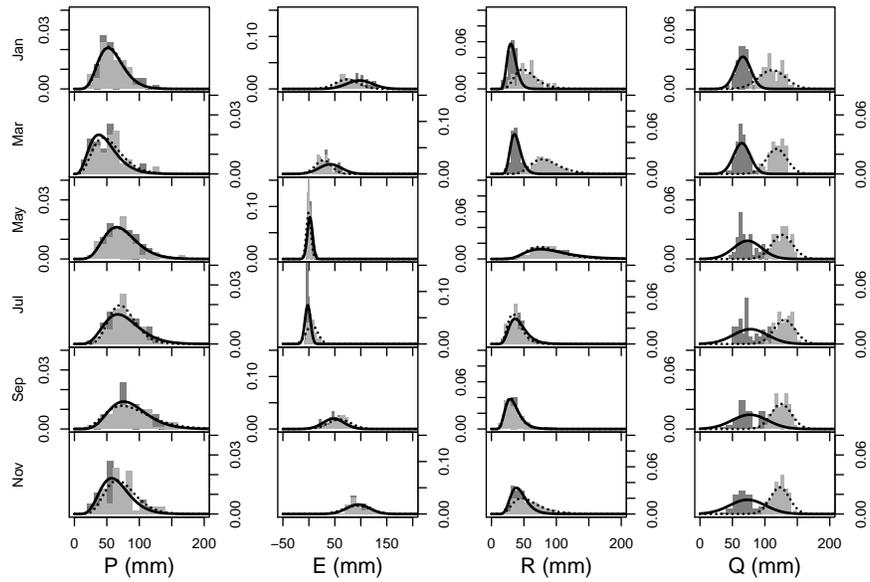}
\caption{Sample of prior distributions for all $\theta$, except diversions ($D$), for every other month starting with January and ending with November.
We constructed prior distributions in such a way to capture seasonal variability, if any, across components. Histograms represent the
available historical data supporting the fitted prior distributions while the lines represent the fitted density functions. Dark gray
histogram bars with solid black density fit lines represent Lake Superior's priors, and light gray bars with dotted density fit lines represent 
Lake Michigan-Huron's priors.}
\label{fig:priors}
\end{figure}

For water balance component prior distributions ($\pi(\theta_{l,t})$), 
we fitted probability density functions to historical monthly data, aggregated by the 12 calendar months (sample
of distributions shown in figure \ref{fig:priors}). 
To calculate prior distribution parameters, we used data for the period 1950 thru 2004 from the 
GLM-HMD and coordinated data sets -- consistently available data from before the analysis period of 2005-2014. 
While we could have used less informative priors, we believe it would have been indefensible to use vague or 
non-informative priors given available data and expert hydrological knowledge.

True values of evaporation ($E$), channel flow ($Q$), and diversion ($D$) variables were given normal prior distributions:

\begin{eqnarray}
    \pi(E_{l,t})         &   \sim   &    \mathsf{N}(\hat{\mu}_{l,E,c(t)}, 0.5\hat{\tau}_{l,E,c(t)})\label{eq:eprior}\\
		\pi(Q_{l,t})         &   \sim   &    \mathsf{N}(\hat{\mu}_{l,Q,c(t)}, 0.5\hat{\tau}_{l,Q,c(t)})\label{eq:qprior}\\
		\pi(D_{l,t})         &   \sim   &    \mathsf{N}(\hat{\mu}_{l,D,c(t)}, \hat{\tau}_{l,D,c(t)})\label{eq:dprior}
\end{eqnarray}

\noindent with means ($\hat{\mu}_{l,*,c(t)}$) and precisions ($\hat{\tau}_{l,*,c(t)}$)  
calculated from the historical data. Because the range of historical values of evaporation and channel outflows is quite narrow for the late 
spring and early summer months of the year, we halved the calculated precisions of the prior probability distributions for 
evaporation and channel outflows to accommodate factors such as climate change \citep[][]{mill:beta:falk:etal:2008}. We do not 
expect empirically-derived prior probability distributions for other, more intrinsically variable water balance components to 
restrict the estimation of their respective posterior probability distributions.

For the true values of non-negative components precipitation and runoff, we fitted, respectively, Gamma 
\cite{husak2007use} and log-normal prior distributions:

\begin{eqnarray}
    \pi(P_{l,t})         &   \sim   &    \mathsf{Gamma}(\psi_{1,l,c(t)}, \psi_{2,l,c(t)})\label{eq:pprior}\\
		\pi(R_{l,t})         &   \sim   &    \mathsf{ln\ N}(\hat{\mu}_{l,ln(R),c(t)}, \hat{\tau}_{l,ln(R),c(t)})\label{eq:rprior}
\end{eqnarray}

For runoff, we calculated the means ($\hat{\mu}_{l,ln(R),c(t)}$) and precisions ($\hat{\tau}_{l,ln(R),c(t)}$) for the prior 
distributions with the natural logarithms of the historical data. 
Gamma distributions for precipitation require non-trivial maximum likelihood shape ($\psi_{1,l,c(t)}$, equation \ref{eq:shape}) 
and rate ($\psi_{2,l,c(t)}$, equation \ref{eq:rate}) parameters 
(per Thom (1958) \citep{thom:1958}) which we calculated from the historical data such that:

\begin{eqnarray}
    \psi_{1,l,c(t)}      &   =   &   \frac{1}{4\phi_{l,c(t)}}\left(1+\sqrt{1+\frac{4\phi_{l,c(t)}}{3}}\right) \label{eq:shape}\\
    \phi_{l,c(t)}        &   =   &   \ln(\hat{\mu}_{l,P,c(t)})-\hat{\mu}_{l,ln(P),c(t)}\notag \\
    \psi_{2,l,c(t)}      &   =   &   \psi_1/\hat{\mu}_{l,P,c(t)} \label{eq:rate}
\end{eqnarray}

\noindent where $\hat{\mu}_{l,P,c(t)}$ and $\hat{\mu}_{l,ln(P),c(t)}$ are the means of the
historical precipitation observations for a given calendar month $c(t)$ and natural logarithms thereof, 
respectively. While channel flow and diversion values are generally non-negative over the course of a 
calendar month, we believe inferences of channel flows and diversions will not be negative, and that
any negative inferences of diversions will be negligible given their scale compared to other water 
balance components. 

We included in supplementary material for this manuscript histograms of all seasonal historical data for each water balance component
and each lake, along with the fitted prior distributions. No variations in $\pi(\theta_{l,t})$ formulations are
examined for this manuscript.

\subsubsection{Likelihoods}\label{sec:lf}

We linked the water balance and independent estimates of water balance components via normal distributions.
Incorporating our beliefs of a data source's potential bias, we established for lake $l$ a relationship between the observed 
value of a component $\theta$ from data source $n$ on month $t$ ($y_{l,\theta,n,t}$) to the true value of $\theta_{l,t}$ via the likelihood:

\begin{eqnarray}
		y_{l,\theta,n,t} & \sim  & \mathsf{N}(\theta_{l,t} + \eta_{l,\theta,n,t}, \tau_{l,\theta,n}), \theta \in [P,E,R,Q,D] \label{eq:clike}
\end{eqnarray}

\noindent with $\eta_{l,\theta,n,t}$ representing observation bias, and component likelihood precision $\tau_{l,\theta,n}$
given a vague $\mathsf{Gamma}(0.1,0.1)$ prior. For non-negative variables precipitation and runoff, we assumed that the non-negative,
asymmetric priors we applied will ensure non-negative inferences of the components' true values, despite the application
of symmetric and potentially negative normal likelihoods. 

We formally considered two alternative structures for bias $\eta_{l,\theta,n,t}$ of data
sources $y_{l,\theta,n,t}$ (equation \ref{eq:clike}). Unlike process error ($\epsilon$), we did not consider an alternative where
$\eta_{l,\theta,n,t} = 0$, as we believe biases are inherent in independent water balance component
estimates, such as precipitation (see Holman et al. (2012), \citep{holman2012improving}). Thus, the first alternative
for $\eta_{l,\theta,n,t}$ we used, following 
the structure of equation \ref{eq:error1} and the prototype model's implementation, strictly tracks seasonal variation:

\begin{eqnarray}
  \eta_{l,\theta,n,t} & = & \eta_{l,\theta,n,c(t)}\notag \\
	\pi(\eta_{l,\theta,n,c(t)}) & \sim  & \mathsf{N}(0,0.01)\label{eq:bias1}
\end{eqnarray}

\noindent The second alternative structure for bias, following the structure of equation \ref{eq:error2}, is hierarchical:

\begin{eqnarray}
\pi(\eta_{l,\theta,n,t})                & \sim  & \mathsf{N}(\eta_{l,\theta,n,c(t)}, \tau_{l,\eta,\theta,n,c(t)})\label{eq:bias2}\\
\pi(\eta_{l,\theta,n,c(t)})             & \sim  & \mathsf{N}(0,0.01)\notag \\
\pi(\tau_{l,\eta,\theta,n,c(t)})        & \sim  & \mathsf{Gamma}(0.05,0.05)\notag
\end{eqnarray}

Lastly, we reflected the \emph{a priori} opinions of regional water resource management authorities regarding the accuracy 
of channel flow ($Q$) and diversion ($D$) estimates through similar bias constructions. Following informal protocols 
for soliciting \emph{a priori} expert opinions \citep{bors:clem:magu:reck:2001, modellingWithStakeHolders}, we found 
that regional water management authorities believe that monthly channel flow data can depart from true channel flows 
by between roughly 180 and 270 cubic meters per second (cms) -- 6 to 9mm of water over the surface of Lake Superior, 
and 4 to 6mm of water over the surface of Lake Michigan-Huron. Departure of diversion estimates from their true values, 
given their smaller magnitudes, are less (see table \ref{tab:data}). In a theoretical 95\% credible interval produced from the previously 
defined $\eta_{l,\theta,n,t}$ (equations \ref{eq:bias1} and \ref{eq:bias2}), water balance component observations may depart from the true 
value of components by about 20 mm (precision = 0.01, standard deviation is thus 10), more than 
double the suggested departure of channel flow and diversion estimates. 

Thus, for $\zeta \in (Q,D)$ in equations below, we modified equations \ref{eq:bias1} and \ref{eq:bias2}, 
increasing the prior precision of the seasonal bias parameter ($\eta_{l,\zeta,n,c(t)}$). We recognize we could have done 
a more specific prior for each estimate of each channel flow or diversion. However, in the interest of minimizing
experimental models to run and testing solely the impact of restricting bias on channel flow and diversion estimates,
we applied a prior precision of 0.25 for the seasonal bias of all estimates of channel flows and diversions (standard deviation
of 2 mm, maximum theoretical 95\% credible interval departure of 4 mm), accounting for a second set of experiments in section \ref{sec:design}. 
Equation \ref{eq:bias1} thus becomes:

\begin{eqnarray}
	\eta_{l,\zeta,n,t} & = & \eta_{l,\zeta,n,c(t)} \label{eq:bias3} \\ 
	\pi(\eta_{l,\zeta,n,c(t)}) & \sim & \mathsf{N}(0,0.25)\notag 
\end{eqnarray}

\noindent and equation \ref{eq:bias2} thus becomes:

\begin{eqnarray}
\eta_{l,\zeta,n,t}            & \sim  & \mathsf{N}(\eta_{l,\zeta,n,c(t)}, \tau_{l,\eta,\zeta,n,c(t)}) \label{eq:bias4}\\
\eta_{l,\zeta,n,c(t)}         & \sim  & \mathsf{N}(0,0.25)\notag \\
\tau_{l,\eta,\zeta,n,c(t)}   & \sim  & \mathsf{Gamma}(0.05,0.05)\notag
\end{eqnarray}

\noindent Structures of $\eta_{l,\theta,n,t}$ represented in equations \ref{eq:bias1}, \ref{eq:bias2}, \ref{eq:bias3}, and \ref{eq:bias4} 
are intended to, like structures of $\epsilon$ (equations \ref{eq:error1} and \ref{eq:error2}), isolate water balance 
component observation uncertainty from uncertainty in the water balance. 
We therefor modeled $\eta_{l,\theta,n,t}$ at seasonal and monthly temporal resolutions, respectively, while $\tau_{l,\theta,n}$
(equation \ref{eq:clike}) is assumed to be constant throughout the analysis period.

We recognize that we utilized three different Gamma distributions as prior distributions for precisions ($\tau$), 
which expressed different beliefs in the variability of particular parameters. $\tau_{l,\Delta{H},w}$ (equation 
\ref{eq:wlobs}) were given $\mathsf{Gamma}(0.01,0.01)$ priors, which have their greatest densities 
towards the lower precision values, reflecting our desire for the model to explore a range of change in storage 
and water balance component values in producing a closed water balance. $\tau_{l,\theta,n}$ (equation \ref{eq:clike}) 
were given $\mathsf{Gamma}(0.1,0.1)$ priors, and have higher densities for higher 
precision values to propagate uncertainty into component observation bias terms $\eta_{l,\theta,n,t}$. 
Lastly, $\tau_{l,\epsilon,c(t)}$, and $\tau_{l,\eta,\theta,n,c(t)}$ (equation \ref{eq:error2} and \ref{eq:bias2} 
respectively) were given $\mathsf{Gamma}(0.05,0.05)$ priors, expressing our belief that monthly process 
errors and observation biases may vary from an unobserved, seasonal cycle at a magnitude less than the variability of 
$y_{\Delta{H},j,w}$ and more than the variability of $y_{l,\theta,n,t}$.

Inferences of $\tau$ and $\eta$ are analyzed and discussed in section \ref{vPrior}.

\subsection{L2SWBM Variations}\label{sec:design}

We formally analyzed combinations of L2SWBM parameter structures, or L2SWBM variations, detailed in sections \ref{sec:wbf} and \ref{sec:wbct}  
using a modified factorial experiment design \citep{montgomery2008design}. Table \ref{tab:design} lists
the 26 models we assessed, incorporating options laid out in sections \ref{sec:wbf} and \ref{sec:wbct}. 
Only two L2SWBMs, PROT and fPROT (for prototype) used the balance computation in equation \ref{eq:pwb} due to
computational expense, differing by the modification to the prior on channel flow and diversion bias 
detailed in equation \ref{eq:bias3}. The prototype L2SWBMs may be compared directly to models (f)01NF and (f)12NF as while they
utilize balance equation \ref{eq:wb}, all other experimental factors are the same as the prototype. The remaining 24 L2SWBMs
vary by rolling window length (1 or 12 months), whether and how they modeled process error (`N'one, `F'ixed, or 
`H'ierarchical), how they modeled water balance component observation bias (`F'ixed or `H'ierarchical), and whether or not
they constrained inference of bias for channel flows and diversions (equation \ref{eq:bias3} or \ref{eq:bias4}, indicated
by prefix `f' for the model ID in the left-most column).

We consider the prototype model, utilizing balance equation \ref{eq:pwb}, to be computationally cost prohibitive for 
temporal and spatial expansion (across all Great Lakes). The number of calculations required per analysis time step 
in the prototype increases as the model iterates through the analysis period.
For our non-prototype experimental models utilizing equation \ref{eq:wb}, however, the number of required calculations per time step is fixed.

\begin{table*}
\centering
\resizebox{0.9\textwidth}{!}{
\begin{tabular}{ c  l  l  l  l }
\hline
\multicolumn{1}{c}{Model ID} & \multicolumn{1}{c}{$w$}  & \multicolumn{1}{c}{$\pi(\epsilon_{l,t})$)} &  \multicolumn{1}{c}{$\pi(\eta_{l,\theta,n,t})$} & $\pi(\eta_{l,\zeta,n,t})$    \\
\hline
PROT  & C (eq. \ref{eq:pwb}) &  None                             & \multicolumn{2}{l}{Fixed (eq. \ref{eq:bias1})}                         \\
01NF  & 1  (eq. \ref{eq:wb}) & (N)one                            & \multicolumn{2}{l}{(F)ixed (eq. \ref{eq:bias1})}                       \\
01NH  & 1  (eq. \ref{eq:wb}) & (N)one                            & \multicolumn{2}{l}{(H)ierarchical (eq. \ref{eq:bias2})}                \\
01FF  & 1  (eq. \ref{eq:wb}) & (F)ixed (eq. \ref{eq:error1})              & \multicolumn{2}{l}{(F)ixed (eq. \ref{eq:bias1})}              \\
01FH  & 1  (eq. \ref{eq:wb}) & (F)ixed (eq. \ref{eq:error1})              & \multicolumn{2}{l}{(H)ierarchical (eq. \ref{eq:bias2})}       \\
01HF  & 1  (eq. \ref{eq:wb}) & (H)ierarchical (eq. \ref{eq:error2})       & \multicolumn{2}{l}{(F)ixed (eq. \ref{eq:bias1})}             \\
01HH  & 1  (eq. \ref{eq:wb}) & (H)ierarchical (eq. \ref{eq:error2})       & \multicolumn{2}{l}{(H)ierarchical (eq. \ref{eq:bias2})}      \\
12NF  & 12 (eq. \ref{eq:wb}) & (N)one                            & \multicolumn{2}{l}{(F)ixed (eq. \ref{eq:bias1})}                      \\
12NH  & 12 (eq. \ref{eq:wb}) & (N)one                            & \multicolumn{2}{l}{(H)ierarchical (eq. \ref{eq:bias2})}               \\
12FF  & 12 (eq. \ref{eq:wb}) & (F)ixed (eq. \ref{eq:error1})              & \multicolumn{2}{l}{(F)ixed (eq. \ref{eq:bias1})}             \\
12FH  & 12 (eq. \ref{eq:wb}) & (F)ixed (eq. \ref{eq:error1})              & \multicolumn{2}{l}{(H)ierarchical (eq. \ref{eq:bias2})}      \\
12HF  & 12 (eq. \ref{eq:wb}) & (H)ierarchical (eq. \ref{eq:error1})       & \multicolumn{2}{l}{(F)ixed (eq. \ref{eq:bias1})}             \\
12HH  & 12 (eq. \ref{eq:wb}) & (H)ierarchical (eq. \ref{eq:error1})       & \multicolumn{2}{l}{(H)ierarchical (eq. \ref{eq:bias2})}      \\
\hline                                                                                                                    
fPROT & C (eq. \ref{eq:pwb}) &  None                             & Fixed (eq. \ref{eq:bias2})           & (eq. \ref{eq:bias3})           \\
f01NF & 1  (eq. \ref{eq:wb}) & (N)one                            & (F)ixed (eq. \ref{eq:bias1})         & (eq. \ref{eq:bias3})           \\
f01NH & 1  (eq. \ref{eq:wb}) & (N)one                            & (H)ierarchical (eq. \ref{eq:bias2})  & (eq. \ref{eq:bias4})           \\
f01FF & 1  (eq. \ref{eq:wb}) & (F)ixed (eq. \ref{eq:error1})              & (F)ixed (eq. \ref{eq:bias1})         & (eq. \ref{eq:bias3})  \\
f01FH & 1  (eq. \ref{eq:wb}) & (F)ixed (eq. \ref{eq:error1})              & (H)ierarchical (eq. \ref{eq:bias2})  & (eq. \ref{eq:bias4})  \\
f01HF & 1  (eq. \ref{eq:wb}) & (H)ierarchical (eq. \ref{eq:error1})       & (F)ixed (eq. \ref{eq:bias1})         & (eq. \ref{eq:bias3}) \\
f01HH & 1  (eq. \ref{eq:wb}) & (H)ierarchical (eq. \ref{eq:error1})       & (H)ierarchical (eq. \ref{eq:bias2}   & (eq. \ref{eq:bias4}) \\
f12NF & 12 (eq. \ref{eq:wb}) & (N)one                            & (F)ixed (eq. \ref{eq:bias1})         & (eq. \ref{eq:bias3})          \\
f12NH & 12 (eq. \ref{eq:wb}) & (N)one                            & (H)ierarchical (eq. \ref{eq:bias2})  & (eq. \ref{eq:bias4})          \\
f12FF & 12 (eq. \ref{eq:wb}) & (F)ixed (eq. \ref{eq:error1})              & (F)ixed (eq. \ref{eq:bias1})         & (eq. \ref{eq:bias3}) \\
f12FH & 12 (eq. \ref{eq:wb}) & (F)ixed (eq. \ref{eq:error1})              & (H)ierarchical (eq. \ref{eq:bias2})  & (eq. \ref{eq:bias4}) \\
f12HF & 12 (eq. \ref{eq:wb}) & (H)ierarchical (eq. \ref{eq:error1})       & (F)ixed (eq. \ref{eq:bias1})         & (eq. \ref{eq:bias3}) \\
f12HH & 12 (eq. \ref{eq:wb}) & (H)ierarchical (eq. \ref{eq:error1})       & (H)ierarchical (eq. \ref{eq:bias2})  & (eq. \ref{eq:bias4}) \\
\hline
\end{tabular}
}
\caption{Summary of our experimental design in which alternative models are configured with
variations in the length of monthly water balance window (used in model inference) $w$, prior probability
distributions for process error $\pi(\epsilon_{l,t})$, and prior probability distribution for data bias $\pi(\eta_{l,\theta,n,t})$.
A window of $C$ indicates the water balance of equation \ref{eq:pwb} was used in the model (see section \ref{sec:wbf}).}
\label{tab:design}
\end{table*}

\subsubsection{Model evaluation}\label{sec:eval}

Experimentation is intended to find a robust L2SWBM formulation that adequately closes the water balance, 
comparable to the prototype, while doing so more efficiently. With our experiment design and 24 non-prototype 
L2SWBMs, we aimed to identify L2SWBM structures that better close the water balance and 
better converge \citep{gelman1992inference} on true values for our parameters of interest. 
We also aimed to identify L2SWBM structures that should not be employed due to inadequate performance
in terms of balance closure or convergence. In fact, some of the proposed L2SWBM structures may be erroneous 
\citep{bulygina2009estimating}, and may be exposed through poor closure or convergence in comparison with 
other L2SWBMs. Water balance closure addresses a need in data to support water resource management decisions, 
while increased L2SWBM efficiency enables exploration of alternative L2SWBMs, which will be critical
when we expand L2SWBMs to include the remaining Laurentian Great Lakes. 

In analyzing water balance closure of the experimental L2SWBMs, we assessed closure 
across all 1, 12, and 60-month periods from 2005 through 2014, instead of closure through the entire 120 
month period of study, which is less practical from an operational standpoint. Posterior predictive distributions 
\citep{gelman2013philosophy, kruschke2013posterior} simulate what observations of a variable could be given all other 
information in a model except the observations. We constructed the posterior predictive distributions for 
observations of change in storage such that:

\begin{eqnarray}
		\tilde{y}_{l,\Delta{H},j,1} & \sim & \mathsf{N}(\Delta{H}_{l,j,1}, \tau_{l,\Delta{H},w})\label{eq:ypp1} \\
		\tilde{y}_{l,\Delta{H},j,12} & \sim & \mathsf{N}(\Delta{H}_{l,j,12}, \tau_{l,\Delta{H},w})\label{eq:ypp12} \\
		\tilde{y}_{l,\Delta{H},j,60} & \sim & \mathsf{N}(\Delta{H}_{l,j,60}, \tau_{l,\Delta{H},w})\label{eq:ypp60} 
\end{eqnarray}


\noindent where $w$ is the rolling window used in an experimental L2SWBM. We used the same change in storage observation precision
learned in the model to generate posterior predictive samples.
Using MCMC samples from distributions \ref{eq:ypp1}, \ref{eq:ypp12}, and \ref{eq:ypp60}, we derived 95\%
credible intervals for, respectively, changes in storage across all 1, 12, and 60-month periods from 2005 through 2014.
We then calculated the corresponding observed changes in storage: $y_{l,\Delta{H},j,1}$, $y_{l,\Delta{H},j,12}$, 
$y_{l,\Delta{H},j,60}$. Hence, an experimental L2SWBM closes the water balance at a rate equal to
the number of observations ($y_{l,\Delta{H},j,1}$, $y_{l,\Delta{H},j,12}$, $y_{l,\Delta{H},j,60}$) that are within the 95\% 
credible intervals of their respective posterior predictive distributions ($\tilde{y}_{l,\Delta{H},j,1}$, 
$\tilde{y}_{l,\Delta{H},j,12}$, $\tilde{y}_{l,\Delta{H},j,60}$), divided by the respective number of observations 
(120, 109, and 61).

We recognize that our L2SWBM evaluation design (table \ref{tab:design}) does not include water balance component 
inference using an L2SWBM with a rolling window ($w$) of 60 months and, furthermore, that a range of different rolling
windows could have been explored for inference in our experiment design.  Regardless, our approach addresses the
question of whether water balance components inferred from an L2SWBM with only a 1-month rolling window 
(which may have the advantage of a relatively short computation time) close the water balance over periods longer 
than one month, or if a longer rolling window (in our case, assessed using a 12-month window) is needed.

For water balance closure assessment, we ran each L2SWBM alternative for $K = 250,000$ MCMC iterations across 
three parallel MCMC chains, and thinned the last 125,000 iterations of each chain (omitting the first 125,000 
iterations as a `burn-in' period) at even intervals such that the resulting chains each have 1,000 
values.  The resulting 3,000 MCMC samples for each parameter then serve as the basis 
for our posterior predictive water balance closure assessment, along with our posterior inferences of other
parameters of interest (e.g. channel flows and diversions). We ran the L2SWBMs on a Windows 7 Professional 
(Microsoft, Redmond, WA, USA) workstation with a 64-bit Intel Core i7-3770 (3.4 GHz) processor with 32 GB 
of RAM. JAGS model code is included in the supplementary material.

Additionally, for each L2SWBM in our experimental design, we analyzed model convergence through
$K = 250,0000$ MCMC iterations by calculating the potential scale reduction factor
(PSRF), also referred to as the Gelman-Rubin convergence statistic or $\hat{R}$ \citep{gelman1992inference},
every 10,000 iterations for all L2SWBM parameters. We, again, at every 10,000 iterations, treated half of the samples
of each chain as burn-in, and thinned the latter half of each chain down to 1,000 samples. We calculated the PSRF using the
\texttt{gelman.diag} function in the R package `CODA' which returned both the median
($\hat{R}_{50}$) and 97.5$\%$ quantile ($\hat{R}_{97.5}$) of the PSRF.  We assessed convergence by testing
whether $\hat{R}_{50}$ and $\hat{R}_{97.5}$ approached (or decreased below) 1.1, per guidance from Gelman and Rubin (1992),
with additional MCMC iterations. Parallel to assessing L2SWBM efficiency in terms of iterations required for convergence, 
we recorded the total time required for each L2SWBM to generate $K = 250,000$ MCMC iterations. The L2SWBM runs we executed 
for monitoring computation time did not include calculations for posterior predictive distributions because they serve as a 
basis for model verification only and would not, we believe, be encoded in a future version employed in routine operations.

In concluding L2SWBM comparison, we calculated the Deviance Information Criteria (DIC, \cite{spiegelhalter2002bayesian}), 
as an MCMC-friendly substitute for the Akaike information criteria (AIC) and Bayesian information 
criteria (BIC). We drew 1000 samples via the \texttt{dic.samples} function in the R package `CODA' using the `pD' penalty 
type and calculated a mean score for each L2SWBM in our experimental design. We then compared the scores to the other tests we perform.

\section{Application of Methodology}

\subsection{Analysis of water balance closure}

Our analysis of 95$\%$ posterior predictive intervals for simulated changes in lake storage across 1, 12, and 60-month periods
 (table \ref{tab:closureResults}) indicates that most L2SWBMs conditioned on changes in storage across a 1-month window close 
the water balance for Lakes Superior and Michigan-Huron in simulations over a 1-month period, but do not effectively close the 
water balance in simulations over 12 and 60-month periods. Water balance component estimates conditioned on a rolling 12-month 
storage window, however, close the water balance for 12- and 60-month periods for Lakes Superior and Michigan-Huron, and come 
close to closing the water balance for both lakes on a 1-month storage window.  For example, 95$\%$ posterior predictive 
intervals for monthly changes in storage derived from 12-month rolling window models included between 91$\%$ and 100$\%$ of the 
observed monthly changes in storage for Lake Superior, and between 79$\%$ and 98$\%$ of observed monthly changes in storage for 
Lake Michigan-Huron.  Similarly, 95$\%$ posterior predictive intervals for 12- and 60-month changes in storage derived from 12-
month rolling window models contained 97$\%$ to 100$\%$ of the observed monthly changes in storage for both Lakes Superior and 
Michigan-Huron. 

Our results show slight improvements in balance closure occur through either a) the introduction alone of an explicit process error 
term or b) the relaxation of prior probability distributions for the bias of data sources via the introduction of a hierarchical structure.
Applying a reduced range of bias in channel flow measurements yielded mixed results in terms of improving balance closure in an L2SWBM.
L2SWBMs that inferred changes in storage over a 1-month window exhibited little impact on 1-month window closure rates post-application, in 
contrast to a maximum 17\% drop (01NH to f01NH) post-application for 12 and 60 month closure rates. 12-month rolling window
models were more robust, with the only decreases in performance occurring with 1-month changes in 
storage on Michigan-Huron \textemdash\ the percentage of observed changes in lake storage within 95\% posterior predictive credible 
intervals dropped between 2\% and 11\% after application.

\begin{figure}
	\centering
		\includegraphics[width=0.9\textwidth]{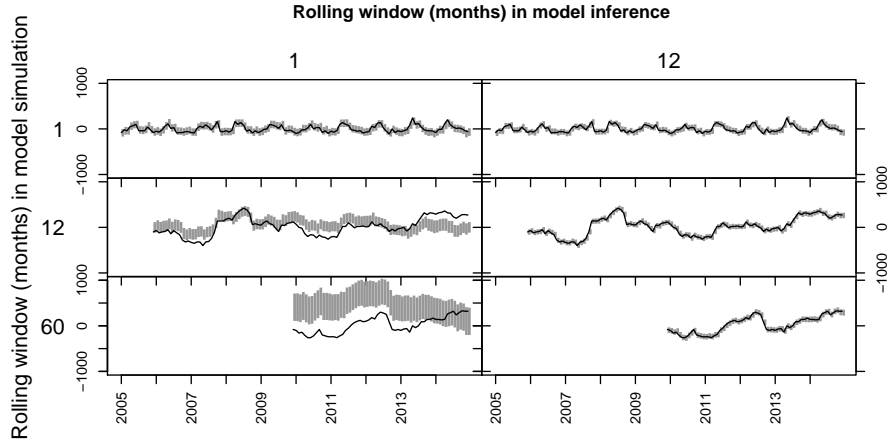}
	\caption{Comparison between observed (black line) and simulated (95$\%$ posterior predictive intervals; grey regions) 
	changes in lake storage (mm over lake surface) across a period of 1, 12, and 60 months (indicated adjacent to left-hand
	vertical axes) for Lake Superior. Simulated changes are based on water balance components inferred using either a 1-month (left column)
	or 12-month (right column) window. Simulation results shown are based on the f01FF and f12FF models (per table 
	\ref{tab:closureResults}). We note that models with a 1-month balance window do not adequately simulate long-term
	changes in storage, indicated by the black line falling outside of the gray regions plotted for those models 
	and simulation windows. Visual comparison for Lake Michigan-Huron is omitted due to redundancy.}
	\label{fig:closureAnalysisGraph}
\end{figure}

A visual inspection of a representative time series (from the f01FF and f12FF models) comparing observed and simulated changes 
in storage over 1, 12, and 60-month periods (figure \ref{fig:closureAnalysisGraph}) underscores the degradation in skill when 
water balance components inferred from an 1-month window L2SWBM are used to simulate changes in lake storage across longer time 
periods.  The visual inspection of the representative time series also indicates that while the percentage of observations 
within the 95$\%$ posterior predictive interval for the f12FF model (right column, figure \ref{fig:closureAnalysisGraph}) 
exceeds 95$\%$ when used to simulate 12- and 60-month cumulative changes in storage, overdispersion does not appear to be a
significant problem. L2SWBMs that infer over narrower windows allow more freedom in the exploration of values for components
in a given month. Due to a lack of information from other months, however, it is difficult to close the balance over longer 
time periods with those models. Larger windows for inference, however, shrink the range of possible
values of water balance components, as the values must close a larger number of $w$ month balance periods.

\begin{table}
\centering
\resizebox{0.9\textwidth}{!}{
\begin{tabular}{ c  c  c  c  c  c  c | c  c  c  c  c  c  c }
\hline
    & \multicolumn{13}{c}{Simulation with rolling window (months) of:}\\
    & \multicolumn{2}{c}{1}
    & \multicolumn{2}{c}{12}
    & \multicolumn{2}{c}{60}
		&
		& \multicolumn{2}{c}{1}
    & \multicolumn{2}{c}{12}
    & \multicolumn{2}{c}{60}\\
\hline										
              Model  & SUP  & MHU  & SUP  & MHU  & SUP  & MHU & Model  & SUP  & MHU  & SUP  & MHU  & SUP  & MHU  \\
\hline \hline NULL   & 98   & 98   & 99   & 100  & 100  & 100 & fNULL  & 98   & 98   & 99   & 100  & 100  & 98   \\ 
\hline \hline 01NF   & 97   & 99   & 27   & 29   & 21   & 15  & f01NF  & 98   & 98   & 24   & 17   & 13   & 7    \\ 
\hline        01NH   & 98   & 100  & 35   & 38   & 21   & 23  & f01NH  & 98   & 99   & 29   & 21   & 13   & 13   \\ 
\hline        01FF   & 97   & 99   & 28   & 50   & 25   & 26  & f01FF  & 98   & 99   & 28   & 39   & 23   & 23   \\ 
\hline        01FH   & 99   & 100  & 39   & 52   & 25   & 34  & f01FH  & 99   & 100  & 35   & 50   & 21   & 30   \\ 
\hline        01HF   & 100  & 100  & 51   & 79   & 30   & 38  & f01HF  & 100  & 100  & 45   & 75   & 25   & 36   \\ 
\hline        01HH   & 100  & 100  & 55   & 74   & 41   & 38  & f01HH  & 100  & 100  & 50   & 76   & 28   & 38   \\ 
\hline        12NF   & 92   & 85   & 99   & 100  & 100  & 97  & f12NF  & 91   & 79   & 100  & 100  & 100  & 97   \\ 
\hline        12NH   & 95   & 84   & 100  & 100  & 100  & 100 & f12NH  & 95   & 79   & 100  & 100  & 100  & 98   \\ 
\hline        12FF   & 98   & 88   & 99   & 100  & 100  & 97  & f12FF  & 98   & 86   & 99   & 100  & 100  & 97   \\ 
\hline        12FH   & 98   & 92   & 100  & 100  & 100  & 100 & f12FH  & 98   & 88   & 100  & 100  & 100  & 100  \\ 
\hline        12HF   & 93   & 92   & 100  & 100  & 100  & 100 & f12HF  & 95   & 88   & 100  & 100  & 100  & 100  \\ 
\hline        12HH   & 98   & 98   & 100  & 100  & 100  & 100 & f12HH  & 100  & 87   & 100  & 100  & 100  & 100  \\ 
\hline
\end{tabular}
}
\caption{Percent (\%) of observed changes in lake storage within 95$\%$ posterior predictive intervals of model-simulated changes
 in storage across 1, 12, and 60 month periods for Lakes Superior (SUP) and Michigan-Huron (MHU) across all experimental models.}
\label{tab:closureResults}
\end{table}

\subsection{Model convergence and computation time}

\begin{figure}
\centering
\includegraphics[width=0.9\textwidth]{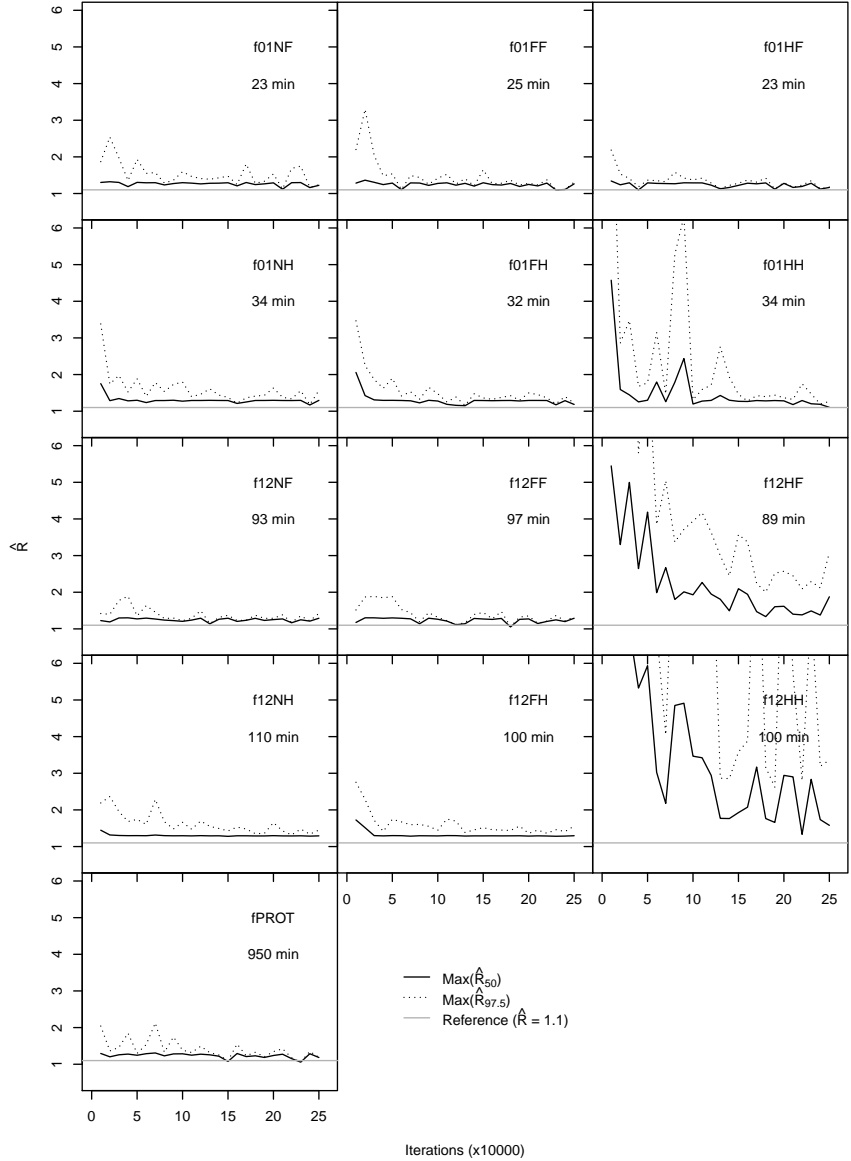}
\caption{Evolution of the maximum value of convergence statistics $\hat{R}_{50}$ and $\hat{R}_{97.5}$ (from all parameters
in a given model) across JAGS sampling iterations for L2SWBMs with informative prior probability
distributions on channel flow and diversion estimate bias. Run times are beneath the L2SWBM label in each panel. We omit figures
illustrating convergence of L2SWBMs without informative prior probability distributions on channel flow and diversion estimate bias 
as those results are redundant.}
\label{fig:convergeAnalysisFlow}
\end{figure}

Results of our convergence analysis (figure \ref{fig:convergeAnalysisFlow}) indicate that most experimental
L2SWBMs approached convergence within 250,000 iterations, but that at least one of the several hundred parameters (up to 4,350 parameters)
in each of our experimental L2SWBMs (represented by the maximum PSRF) did not fully converge.  
We found, for under half the models, just two or tree of the parameters in each model have a PSRF above 1.1. Models 
with a hierarchical data source bias structure, or a combination of 1) a 12-month rolling window
for inference of changes in storage and 2) hierarchical process error and bias structure, had significantly more than three parameters not converge. 
We found it most common, with L2SWBMs that nearly converged completely, for two or three of the water balance component observation precision(s) 
$\tau_{l,\eta,\theta,n,c(t)}$ to be the parameter(s) that did not converge. 
For further reading on analyzing the convergence of a model with many variables, see Brooks and Gelman (1998)\citep{brooks1998general}. 
We omit a figure illustrating convergence of models without informative 
prior probability distributions on channel flow and diversion estimate bias as the results are redundant.

We also find that the time to run an L2SWBM (to 250,000 iterations) with a 12-month rolling inference window (also 
figure \ref{fig:convergeAnalysisFlow}) is roughly 14 hours less than 
the time required for the prototype L2SWBM, but roughly three times longer than the time to run a model with a 1-month inference
 window. The time required to run L2SWBMs with different process error structures are about equal, while implementing a hierarchical structure
for data source bias adds up to 10 minutes compared with a fixed structure. These run times are dependent on the technical 
specifications of the computer used, as well as other
 applications being run simultaneously on the same computer. Run times will, therefore, likely differ across different computational 
environments.

\subsection{DIC results}

\begin{table}
\centering
\resizebox{0.75\textwidth}{!}{
\begin{tabular}{ c c | c c | c c | c c }
\hline
Model  & DIC      & Model &  DIC      & Model  & DIC      & Model &  DIC      \\
\hline \hline
PROT   & 12665.06 & fPROT &  12649.40 &        &          &       &            \\
\hline        
01NF   & 11992.45 & f01NF &  12142.76 & 12NF   & 12582.83 & f12NF &  12733.59  \\
01NH   & 10271.64 & f01NH &  10031.27 & 12NH   &  10831.3 & f12NH &  10595.51  \\
01FF   & 12347.55 & f01FF &  12118.43 & 12FF   & 12558.21 & f12FF &  12507.05  \\
01FH   &  9870.43 & f01FH &   9814.96 & 12FH   & 10622.67 & f12FH &  10579.13  \\
01HF   & 12323.85 & f01HF &  12031.26 & 12HF   & 10824.23 & f12HF &  11221.91  \\
01HH   & 10124.29 & f01HH &  10263.95 & 12HH   &  9143.96 & f12HH &   8237.85  \\
\hline
\end{tabular}
}
\caption{DIC scores for our experimental L2SWBMs using the pD penalty.}\label{tab:dic}
\end{table}

Table \ref{tab:dic} shows the DIC scores for our experimental L2SWBMs. Model f12HH is favored over the rest 
of the experimental L2SWBMs based on the presented DIC scores. Interestingly, the two L2SWBMs that converge
well and satisfactorily close the water balance -- (f)12NF and (f)12FF -- received comparable scores to the
(f)PROT models, with only f12NF receiving a slightly worse score. We note that the DIC scores shown 
exhibit how DIC are not invariant to reparameterization \citep{spiegelhalter2014deviance}. In fact, a 
change in L2SWBM structure results in roughly a 1,000 to 4,000 DIC point difference.

\subsection{Incorporating expert opinions on channel flow bias}

\begin{figure}
\centering
\includegraphics[angle=270, width=1\textwidth]{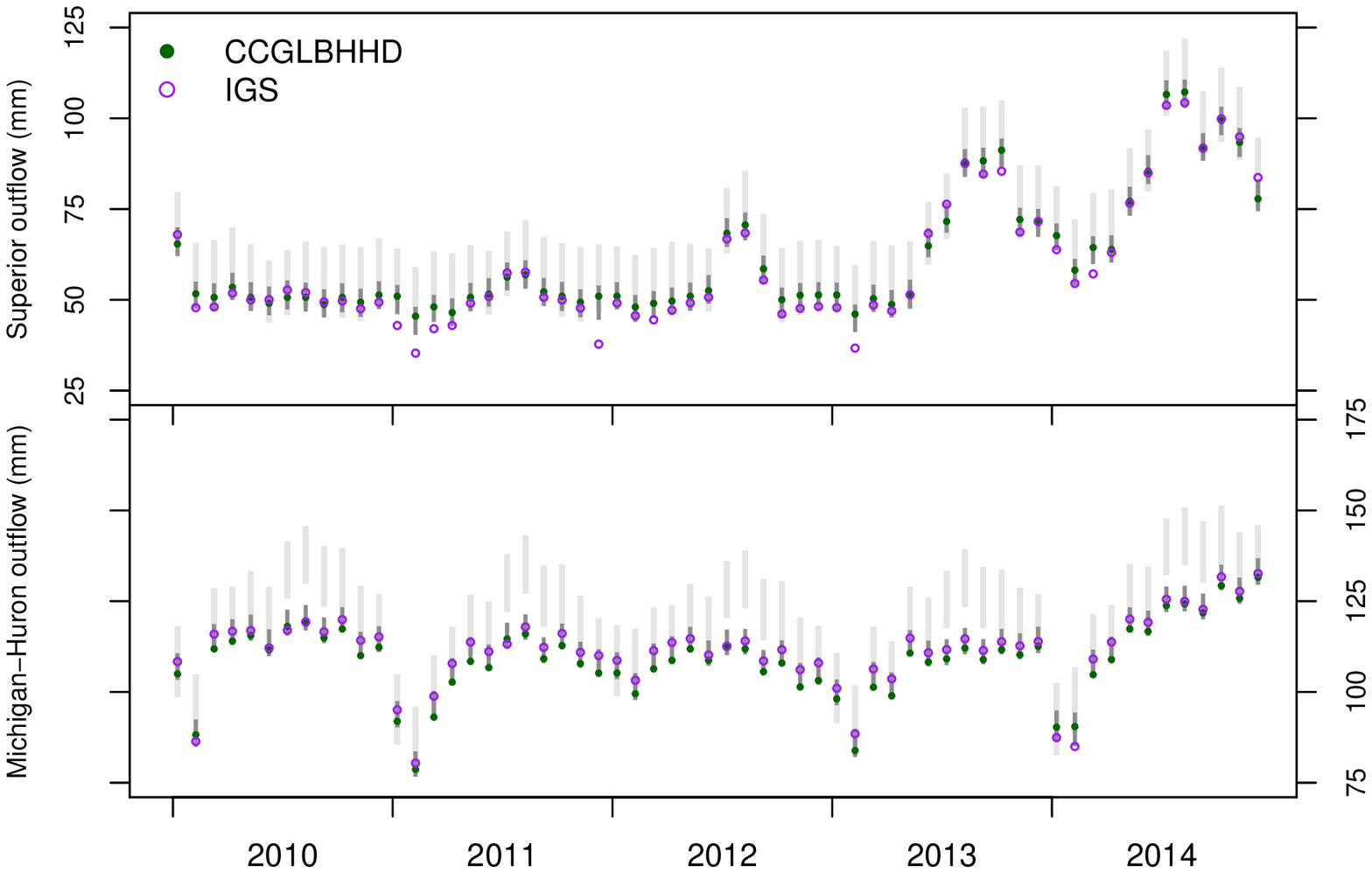}
\caption{Representative estimates (95$\%$ credible intervals) of flow (in mm over the respective lake surface) through the connecting channels
 from Lake Superior (i.e.\ St.\ Marys River) and from Lake Michigan-Huron (i.e.\ St.\ Clair River) from the prototype L2SWBM 
(light grey bars) and experimental L2SWBM f12FF (dark grey bars).  Historical flow measurements are represented by solid dots (CCGLBHHD) and 
circles (IGS).}
\label{fig:flowCompare}
\end{figure}

Reducing the \emph{a priori} range of potential bias in channel flow measurements significantly reduced the uncertainty and 
central tendency of our inferred channel flow estimates (figure \ref{fig:flowCompare}) relative to the prototype model without significantly 
impacting variability and bias in the other inferred water balance components (table \ref{tab:perCompareResults}). For example, we
find that uncertainty, or the standard deviation, in a sample of inferred posterior distributions for precipitation, evaporation, and runoff 
increased by no more than 2 mm from the prototype model to the f12FF and f12NF models. 

\begin{table}
\centering
\resizebox{1.0\textwidth}{!}{
\begin{tabular}{ c | c c c c | c c c c | c c c c }
\hline		
    & \multicolumn{4}{c}{PROT}
    & \multicolumn{4}{c}{f12NF}
    & \multicolumn{4}{c}{f12FF}\\
\hline										
Parameter       
	& Mean & S.D. & 2.5\% & 97.5\% 
	& Mean & S.D. & 2.5\% & 97.5\% 
	& Mean & S.D. & 2.5\% & 97.5\% \\
\hline \hline
\hline $P_{SUP,10}$   & 121.04 &  7.68 & 106.52 & 136.33 & 119.77 &  8.35 & 103.41 & 136.10 & 124.91 &  8.40 & 109.03 & 141.89 \\
\hline $E_{SUP,10}$   &  69.52 &  5.98 &  57.92 &  81.27 &  70.32 &  6.11 &  58.55 &  82.17 &  66.69 &  6.18 &  54.76 &  78.52 \\
\hline $R_{SUP,10}$   &  50.52 &  4.73 &  41.29 &  60.57 &  50.30 &  5.29 &  40.15 &  60.68 &  52.04 &  5.09 &  42.06 &  62.26 \\
\hline $Q_{SUP,10}$   &  66.66 &  4.82 &  57.01 &  75.85 &  61.97 &  1.80 &  58.51 &  65.53 &  61.91 &  1.84 &  58.33 &  65.58 \\
\hline                 
\hline $P_{MHU,10}$   &  24.59 &  6.14 &  13.00 &  37.15 &  28.08 &  7.17 &  15.14 &  42.57 &  31.68 &  7.15 &  18.34 &  46.70 \\
\hline $E_{MHU,10}$   &  91.58 &  5.80 &  80.49 & 102.90 &  88.62 &  6.44 &  76.22 & 101.17 &  84.90 &  6.31 &  72.91 &  97.52 \\
\hline $R_{MHU,10}$   &  19.97 &  3.47 &  13.51 &  27.28 &  21.64 &  3.96 &  14.56 &  29.96 &  23.35 &  4.13 &  15.60 &  31.66 \\
\hline $Q_{MHU,10}$   & 129.83 &  3.89 & 122.55 & 137.63 & 118.56 &  1.60 & 115.43 & 121.65 & 118.31 &  1.63 & 115.15 & 121.62 \\
\hline \hline          
\hline $P_{SUP,101}$  & 109.59 &  7.57 &  94.81 & 124.36 &  95.58 &  8.02 &  80.25 & 111.27 &  98.99 &  8.00 &  82.99 & 114.95 \\
\hline $E_{SUP,101}$  &   1.31 &  3.34 &  -5.37 &   7.75 &   2.80 &  3.28 &  -3.40 &   9.36 &   2.42 &  3.31 &  -3.90 &   9.05 \\
\hline $R_{SUP,101}$  & 178.75 &  6.32 & 165.44 & 190.24 & 166.60 &  7.85 & 150.24 & 180.91 & 171.56 &  7.54 & 156.37 & 185.58 \\
\hline $Q_{SUP,101}$  &  56.61 &  4.28 &  48.42 &  65.32 &  51.68 &  1.71 &  48.36 &  55.08 &  51.53 &  1.74 &  48.15 &  54.96 \\
\hline                 
\hline $P_{MHU,101}$  &  90.60 &  6.95 &  76.64 & 104.32 &  81.48 &  7.40 &  67.22 &  95.56 &  85.29 &  7.63 &  70.69 & 100.32 \\
\hline $E_{MHU,101}$  &  -0.66 &  3.01 &  -6.54 &   5.32 &   0.21 &  3.06 &  -5.94 &   6.16 &  -0.16 &  3.12 &  -6.28 &   6.08 \\
\hline $R_{MHU,101}$  & 132.20 &  4.91 & 122.90 & 141.78 & 126.54 &  5.17 & 116.79 & 136.63 & 129.92 &  5.44 & 119.40 & 140.39 \\
\hline $Q_{MHU,101}$  & 120.16 &  3.87 & 113.07 & 128.10 & 113.73 &  1.54 & 110.75 & 116.85 & 113.46 &  1.53 & 110.44 & 116.46 \\
\hline \hline          
\hline $P_{SUP,113}$  &  85.72 &  8.02 &  70.16 & 101.75 &  77.04 &  8.10 &  60.91 &  92.57 &  80.12 &  8.42 &  63.93 &  96.47 \\
\hline $E_{SUP,113}$  &  -4.74 &  3.47 & -11.49 &   2.05 &  -3.98 &  3.34 & -10.31 &   2.90 &  -4.28 &  3.31 & -10.69 &   2.43 \\
\hline $R_{SUP,113}$  & 165.55 & 11.63 & 142.87 & 188.04 & 148.18 & 12.49 & 124.26 & 172.95 & 153.13 & 12.95 & 128.21 & 178.71 \\
\hline $Q_{SUP,113}$  &  82.30 &  4.25 &  74.29 &  90.90 &  77.33 &  1.67 &  73.97 &  80.63 &  77.21 &  1.72 &  73.72 &  80.60 \\
\hline                 
\hline $P_{MHU,113}$  &  85.26 &  7.73 &  70.19 & 100.71 &  80.70 &  8.21 &  64.92 &  97.11 &  84.11 &  8.20 &  67.99 & 100.04 \\
\hline $E_{MHU,113}$  &  -6.81 &  3.16 & -12.93 &  -0.44 &  -6.72 &  3.08 & -12.51 &  -0.64 &  -6.81 &  3.14 & -12.80 &  -0.79 \\
\hline $R_{MHU,113}$  & 134.12 & 10.90 & 113.24 & 156.45 & 129.10 & 11.43 & 107.10 & 152.20 & 133.63 & 11.38 & 111.09 & 156.16 \\
\hline $Q_{MHU,113}$  & 126.45 &  3.84 & 119.27 & 134.40 & 119.93 &  1.51 & 117.00 & 122.89 & 119.72 &  1.50 & 116.86 & 122.74 \\
\hline
\end{tabular}
}
\caption{Monte Carlo statistics -- mean, standard deviation (S.D.), 2.5\% and 97.5\% quantiles -- for select inferred 
water balance component parameters (in millimeters). We present
the inferred parameters for October 2005 ($t = 10$), May 2013 ($t = 101$), and May 2014 ($t = 113$).}
\label{tab:perCompareResults}
\end{table}

\section{Conclusion and Discussion}

\subsection{L2SWBM selection}

Our experiment indicates that month-to-month water balance inference, or inference using only a one month rolling window, does not close the balance
over time periods of 12 months or greater. As also exposed through our experiment design, MCMC convergence is greatly hindered, if not prevented, given a more complex 
hierarchical process error or component observation bias structure. DIC scores favored the more complex, non-convergent L2SWBMs, but we believe that is an artifact of
DIC's sensitivity to reparameterization. Therefore, two model options can be recommended: f12NF and f12FF. Both options
have a 12-month rolling water balance window, a fixed data source bias structure as in the prototype model, and constrained priors on bias for channel
flow and diversion estimates. Whether or not process error is explicitly estimated on a seasonal basis is the difference between the two options, where the advantage
in the former, or f12FF, is 7\% more months with a closed water balance on a monthly basis. Additionally, both models required roughly 1.5 hours to
compute 250,000 MCMC iterations -- compared to 16 hours for the prototype model -- and closed the water balance at a rate comparable to the prototype model
over 1, 12, and 60 month periods, incorporating current opinions of regional water management authorities. These results give 
water resource managers and analysts flexibility in choosing to estimate process error, which may estimate the collective impact of
groundwater fluxes, isostatic rebound, thermal expansion, and other unmeasured phenomena.

\subsection{Impact of vague priors on posterior inferences of bias and precision}\label{vPrior}

\begin{figure}
	\centering
		\includegraphics[width=0.8\textwidth]{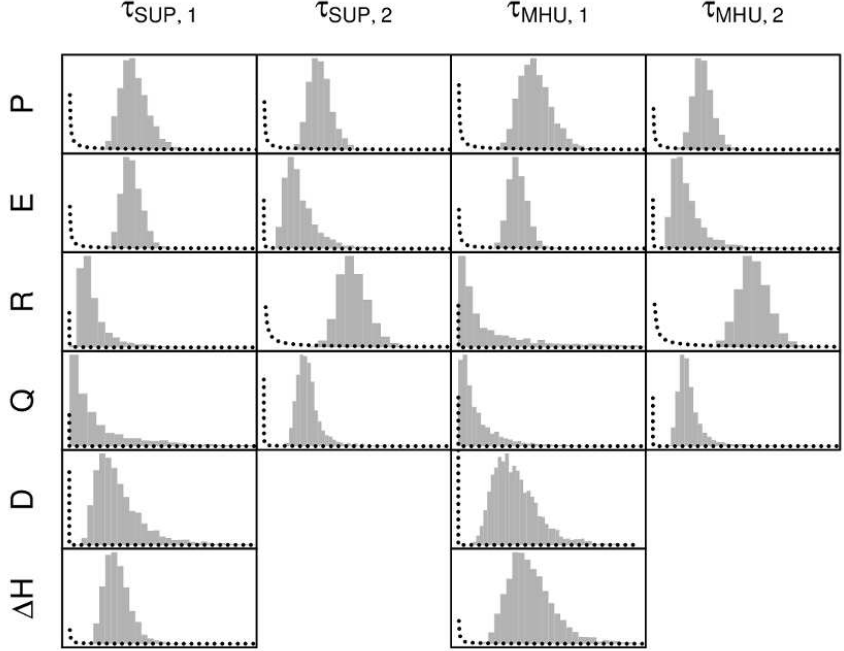}
	\caption{Comparison of vague Gamma priors (dotted lines) and inferred posterior histograms (gray bars)
	for the precision of input estimates for Lakes Superior and Michigan-Huron's water balance parameters. Plots
	utilize representative MCMC samples drawn for model f12FF. Rows represent individual balance parameters and columns indicate
	which lake and input estimate for $\tau$ is plotted. Subscript $\theta$ for column labels is omitted. Priors
	were non-informative for approximately $16/20$ or 80\% of precision parameters.}
	\label{fig:vaguePriorImpactPrecs}
\end{figure}

We found that vague $\mathsf{Gamma}(0.01,0.01)$ and $\mathsf{Gamma}(0.1,0.1)$ priors we applied for 
the precision of observed changes in storage ($\tau_{l,\Delta{H},w}$) and independent water balance component estimates' ($\tau_{l,\theta,n}$) 
were, for the most part, non-informative in our selected models. Figure \ref{fig:vaguePriorImpactPrecs} shows that, for more than half
of the $\tau$ parameters in model f12FF, the MCMC samples of the posterior distribution appear approximately normal in great
contrast to the prior Gamma distribution applied. Table \ref{tab:precSummary} further supports the vague priors as non-informative,
listing the inferred precisions as standard deviations. None of the inferred precisions are, on average, significantly low.
Notably, LBRM estimates were inferred to be more imprecise compared to other 
water balance component estimates, and inferred precisions for ARM estimates were sharply higher, possibly highlighting
better accuracy of stream-gauge based over model-based estimates of runoff into large lakes. Additionally, constraining bias on channel flow and diversion
estimates did not result in significantly lower inferred precisions for the same estimates, comparing the prototype and experimental L2SWBMs.

\begin{figure}
	\centering
		\includegraphics[width=1.0\textwidth]{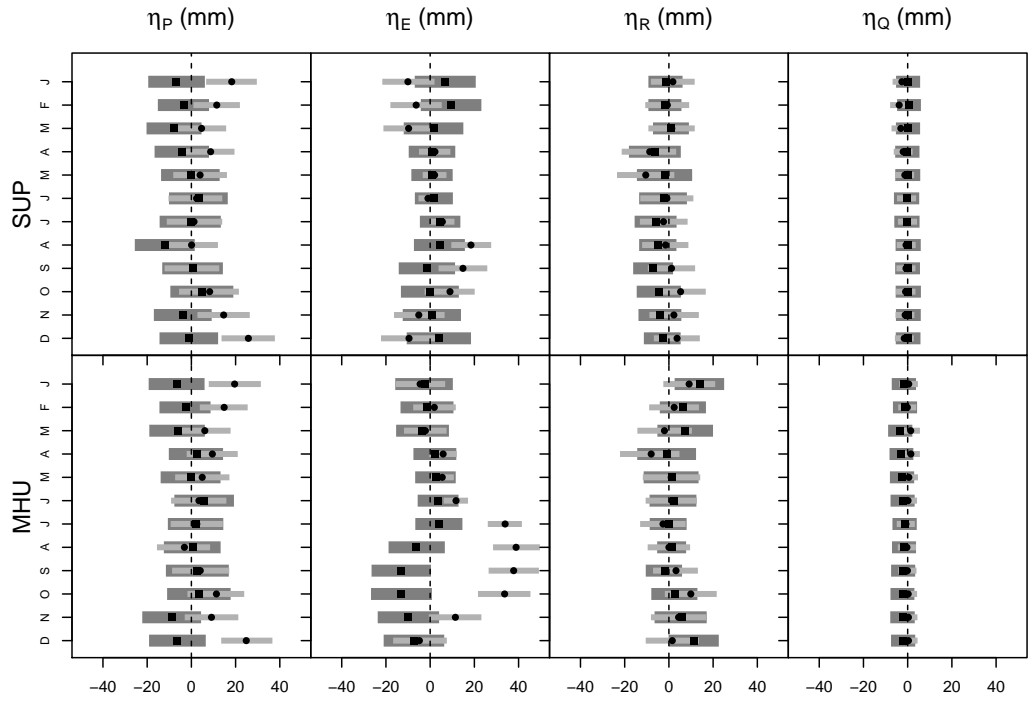}
	\caption{Inferred biases via MCMC samples for model f12FF of independent water balance estimates for (respective of the
	column labels at the top of the figure) precipitation, 
	evaporation, runoff, and channel outflow for Lakes Superior and Michigan-Huron. Each tick along the vertical axes
	represents a calendar month, while the bias value is represented along the horizontal axes. Dark gray bars represent
	the biases' inferred 95\% credible interval for the first set ($n = 1$) of independent estimates for each component, while light 
	gray bars represent the biases' inferred 95\% credible interval	for the second set ($n = 2$) of independent estimates. Black
	squares correspond to the mean inferred bias for the first set ($n = 1$) of independent estimates for each component, while
	black dots correspond to the mean inferred bias for the second set ($n = 2$) of independent estimates.}
	\label{fig:vaguePriorImpactBiases}
\end{figure}

Inferred seasonal biases for independent water balance component estimates $\eta_{l,\theta,n,c(t)}$ were not strongly influenced by their
vague $\mathsf{N}(0,0.01)$ priors (figure \ref{fig:vaguePriorImpactBiases}). While data support negligible bias over the late spring and early summer for
estimates of all components of the water balance, strong biases were inferred for the cooler fall, winter, and spring months. Many factors
may explain the inferred biases, including inaccurate, land-based precipitation measurements in freezing or snowy conditions, poor representation
of lake thermodynamics for evaporation estimates, frozen streams interfering with the accurate measurement of surface runoff into the lakes,
and (despite constrained bias priors for $\eta_{l,\zeta,n,c(t)}$) a lack of compensation for ice jams in channel flow measurements.

\begin{table}
\centering
\resizebox{0.9\textwidth}{!}{
\begin{tabular}{ c | c c c | c c c | c c c }
\hline		
    & \multicolumn{3}{c}{PROT}
    & \multicolumn{3}{c}{f12NF}
    & \multicolumn{3}{c}{f12FF}\\
\hline										
Parameter       & Mean & 2.5\% & 97.5\% & Mean & 2.5\% & 97.5\% & Mean & 2.5\% & 97.5\% \\
\hline \hline
\hline $\sigma_{SUP,P,1}$         &  9.90 & 12.09 &  8.19 &  9.60 & 11.82 & 7.89 &  9.66 & 12.02 &  7.90 \\
\hline $\sigma_{SUP,P,2}$         &  9.99 & 12.16 &  8.27 & 10.32 & 12.65 & 8.61 & 10.43 & 12.61 &  8.72 \\
\hline $\sigma_{SUP,E,1}$         &  9.54 & 11.32 &  8.14 &  9.80 & 11.53 & 8.47 &  9.84 & 11.60 &  8.48 \\
\hline $\sigma_{SUP,E,2}$         &  4.67 &  6.65 &  3.18 &  4.02 &  6.11 & 2.62 &  4.01 &  6.06 &  2.64 \\
\hline $\sigma_{SUP,R,1}$         &  1.85 &  6.10 &  0.65 &  2.36 &  6.59 & 1.08 &  2.81 &  7.00 &  1.88 \\
\hline $\sigma_{SUP,R,2}$         & 14.57 & 17.20 & 12.54 & 14.17 & 16.82 &12.09 & 13.90 & 16.44 & 11.89 \\
\hline $\sigma_{SUP,Q,1}$         &  0.50 &  2.31 &  0.22 &  0.49 &  2.13 & 0.22 &  0.50 &  2.04 &  0.23 \\
\hline $\sigma_{SUP,Q,2}$         &  2.76 &  3.61 &  1.99 &  2.87 &  3.72 & 2.26 &  2.94 &  3.72 &  2.26 \\
\hline $\sigma_{SUP,D,1}$         &  0.78 &  1.25 &  0.49 &  0.76 &  1.22 & 0.47 &  0.77 &  1.24 &  0.49 \\
\hline $\sigma_{SUP,\Delta{H},w}$ & 10.06 & 12.94 &  8.06 & 11.17 & 14.60 & 8.71 & 11.18 & 14.47 &  8.76 \\
\hline                    
\hline $\sigma_{MHU,P,1}$         &  8.84 & 10.91 &  7.25 &  8.81 & 10.92 & 7.25 &  8.83 & 10.97 &  7.25 \\
\hline $\sigma_{MHU,P,2}$         & 10.56 & 12.82 &  8.80 & 10.94 & 13.33 & 9.17 & 11.04 & 13.48 &  9.23 \\
\hline $\sigma_{MHU,E,1}$         & 10.06 & 11.81 &  8.64 & 10.04 & 11.82 & 8.67 & 10.06 & 11.85 &  8.70 \\
\hline $\sigma_{MHU,E,2}$         &  3.23 &  6.14 &  2.10 &  3.42 &  6.27 & 2.17 &  3.71 &  6.36 &  2.29 \\
\hline $\sigma_{MHU,R,1}$         &  0.67 &  3.89 &  0.26 &  0.66 &  4.19 & 0.27 &  0.69 &  4.15 &  0.27 \\
\hline $\sigma_{MHU,R,2}$         & 13.60 & 15.91 & 11.91 & 13.34 & 15.66 &11.61 & 13.25 & 15.50 & 11.49 \\
\hline $\sigma_{MHU,Q,1}$         &  0.47 &  1.59 &  0.22 &  0.47 &  1.64 & 0.21 &  0.48 &  1.66 &  0.22 \\
\hline $\sigma_{MHU,Q,2}$         &  1.41 &  2.06 &  0.82 &  1.29 &  2.02 & 0.70 &  1.36 &  2.03 &  0.76 \\
\hline $\sigma_{MHU,D,1}$         &  0.18 &  0.27 &  0.12 &  0.18 &  0.27 & 0.12 &  0.18 &  0.27 &  0.13 \\
\hline $\sigma_{MHU,\Delta{H},w}$ &  8.62 & 11.28 &  6.77 &  9.02 & 12.57 & 6.63 &  9.14 & 12.65 &  6.81 \\
\hline
\end{tabular}
}
\caption{Monte Carlo statistics for the inferred precisions from L2SWBMs PROT, f12NF, and f12FF, shown as standard deviations (in millimeters), 
with the 2.5 percentile being the high part of the 95\% credible interval, and the 97.5 percentile being 
the low part of the 95\% credible interval.}
\label{tab:precSummary}
\end{table}

\subsection{Future work}

The experiment described in this paper led to the evolution of an L2SWBM for the Laurentian Great Lakes, and has the potential 
to support the development of similar L2SWBMs for other large lake systems around the world. The experiment design can be modified 
to experiment with other inference windows and error structures. Expansion of the new model (or models) 
to Lakes St.\ Clair, Erie, and Ontario, as well as back in time to 1950, is expected to be non-trivial. The number of parameters 
to estimate and required computation time will increase, and there will be other factors to consider, such as water flow through
the Huron-Erie Corridor, and meteorological station and stream gauge availability over time. As the models are tested and improved,
possibly with different model structures, we expect that the resulting water balance component estimates will be employed not only 
by water management authorities, but will be distributed to the public as well through (among other interfaces) the NOAA Great Lakes 
Dashboard Project \citep{dashPaperOne, dashPaperTwo, dashPaperThree}.

\section{Supplementary material}

Supplementary material related to this article can be found with the on-line version of this journal article.
Plots of time series of inferred water balance components from the prototype, model f12NF, and model f12FF are included.
Additionally, compressed folders containing Superior and Michigan-Huron L2SWBM code and data may be downloaded from
https://www.glerl.noaa.gov/data/WaterBalanceModel/. Necessary software packages are available for all major computing
platforms (e.g. Windows, Mac, and Linux).

\section{Acknowledgements}

The authors thank Song Qian, Yves Atchade, Kerby Shedden, Edward Ionides, Vincent Fortin,
Bryan Tolson, and Craig Stow for helpful discussions on Bayesian inference and alternative
formulations of our water balance model. Jacob Bruxer, Frank Seglenieks, Tim Hunter, Tim Calappi and
Lauren Fry provided expert opinions and water balance data. Nicole Rice provided graphical and editorial support.
Funding was provided by the International Joint Commission (IJC) International Watersheds
Initiative (IWI) to NOAA and the Cooperative Institute for Great Lakes Research (CIGLR)
through a NOAA Cooperative Agreement with the University of Michigan (NA12OAR4320071); many thanks
to Wendy Leger and Mike Shantz.
The use of product names, commercial and otherwise, in this paper does not imply endorsement
by NOAA, NOAA-GLERL, CIGLR, or any other contributing agency or organization.  This is
NOAA-GLERL contribution number XXXX and CIGLR contribution number ZZZZ.

A preprint of this article is available. See \citep{2017arXiv171010161S}.

\bibliographystyle{tfs}

\begin{thebibliography}{10}
\providecommand{\MR}{\relax\unskip\space MR }
\providecommand{\url}[1]{\normalfont{#1}}
\providecommand{\urlprefix}{Available at }

\bibitem{ahrestani2013importance}
F.S. Ahrestani, M. Hebblewhite, and E. Post, \emph{The importance of
  observation versus process error in analyses of global ungulate populations},
  Scientific reports 3 (2013), p. 3125.

\bibitem{armero2008bayesian}
C. Armero, A. Lopez-Quilez, and R. Lopez-Sanchez, \emph{Bayesian assessment of
  times to diagnosis in breast cancer screening}, Journal of Applied Statistics
  35 (2008), pp. 997--1009.

\bibitem{arnell1999simple}
N.W. Arnell, \emph{A simple water balance model for the simulation of
  streamflow over a large geographic domain}, Journal of Hydrology 217 (1999),
  pp. 314--335.

\bibitem{arnold1998large}
J.G. Arnold, R. Srinivasan, R.S. Muttiah, and J.R. Williams, \emph{Large area
  hydrologic modeling and assessment part {I}: Model development} (1998).

\bibitem{bates2001markov}
B.C. Bates and E.P. Campbell, \emph{A {M}arkov chain {M}onte {C}arlo scheme for
  parameter estimation and inference in conceptual rainfall-runoff modeling},
  Water resources research 37 (2001), pp. 937--947.

\bibitem{benke2008parameter}
K.K. Benke, K.E. Lowell, and A.J. Hamilton, \emph{Parameter uncertainty,
  sensitivity analysis and prediction error in a water-balance hydrological
  model}, Mathematical and Computer Modelling 47 (2008), pp. 1134--1149.

\bibitem{blangiardo2014evidence}
M. Blangiardo and G. Baio, \emph{Evidence of bias in the eurovision song
  contest: modelling the votes using bayesian hierarchical models}, Journal of
  Applied Statistics 41 (2014), pp. 2312--2322.

\bibitem{bors:clem:magu:reck:2001}
M.E. Borsuk, R. Clemen, L. Maguire, and K.H. Reckhow, \emph{{Stakeholder values
  and scientific modeling in the Neuse River watershed}}, Group Decision and
  Negotiation 10 (2001), pp. 355--373.

\bibitem{boughton2004australian}
W. Boughton, \emph{The {A}ustralian water balance model}, Environmental
  Modelling \& Software 19 (2004), pp. 943--956.

\bibitem{brooks1998general}
S.P. Brooks and A. Gelman, \emph{General methods for monitoring convergence of
  iterative simulations}, Journal of Computational and Graphical Statistics 7
  (1998), pp. 434--455.

\bibitem{bulygina2009estimating}
N. Bulygina and H. Gupta, \emph{Estimating the uncertain mathematical structure
  of a water balance model via bayesian data assimilation}, Water Resources
  Research 45 (2009).

\bibitem{ccghbhhd:1977}
 {CCGLBHHD}, \emph{{Coordinated Great Lakes physical data}}, Tech. {R}ep.,
  {Coordinating Committee on Great Lakes Basic Hydraulic and Hydrologic Data},
  1977. \urlprefix\url{www.lre.usace.army.mil/}.

\bibitem{dashPaperTwo}
A.H. Clites, J.P. Smith, T.S. Hunter, and A.D. Gronewold, \emph{Visualizing
  relationships between hydrology, climate, and water level fluctuations on
  {E}arth's largest system of lakes}, Journal of Great Lakes Research 40
  (2014), pp. 807--811.

\bibitem{croley1989verifiable}
T.E. Croley, \emph{Verifiable evaporation modeling on the laurentian great
  lakes}, Water Resources Research 25 (1989), pp. 781--792.

\bibitem{croley1992long}
T.E. Croley, \emph{Long-term heat storage in the great lakes}, Water resources
  research 28 (1992), pp. 69--81.

\bibitem{croley1985resolving}
T.E. Croley and H.C. Hartmann, \emph{Resolving thiessen polygons}, Journal of
  Hydrology 76 (1985), pp. 363--379.

\bibitem{croley2005distributed}
T.E. Croley and C. He, \emph{Distributed-parameter large basin runoff model.
  {I}: Model development}, Journal of Hydrologic Engineering 10 (2005), pp.
  173--181.

\bibitem{crow2008monitoring}
W.T. Crow, W.P. Kustas, and J.H. Prueger, \emph{Monitoring root-zone soil
  moisture through the assimilation of a thermal remote sensing-based soil
  moisture proxy into a water balance model}, Remote Sensing of Environment 112
  (2008), pp. 1268--1281.

\bibitem{deacu2012predicting}
D. Deacu, V. Fortin, E. Klyszejko, C. Spence, and P.D. Blanken,
  \emph{Predicting the net basin supply to the {G}reat {L}akes with a
  hydrometeorological model}, Journal of Hydrometeorology 13 (2012), pp.
  1739--1759.

\bibitem{dorner2007multi}
S. Dorner, J. Shi, and D. Swayne, \emph{Multi-objective modelling and decision
  support using a {B}ayesian network approximation to a non-point source
  pollution model}, Environmental Modelling \& Software 22 (2007), pp.
  211--222.

\bibitem{eberly2000identifiability}
L.E. Eberly, B.P. Carlin, \emph{et~al.}, \emph{Identifiability and convergence
  issues for {M}arkov chain {M}onte {C}arlo fitting of spatial models},
  Statistics in medicine 19 (2000), pp. 2279--2294.

\bibitem{engeland2002bayesian}
K. Engeland and L. Gottschalk, \emph{Bayesian estimation of parameters in a
  regional hydrological model}, Hydrology and Earth System Sciences Discussions
  6 (2002), pp. 883--898.

\bibitem{engeland2005assessing}
K. Engeland, C.Y. Xu, and L. Gottschalk, \emph{Assessing uncertainties in a
  conceptual water balance model using bayesian methodology/estimation
  bay{\'e}sienne des incertitudes au sein d’une mod{\'e}lisation conceptuelle
  de bilan hydrologique}, Hydrological Sciences Journal 50 (2005).

\bibitem{fry2013identifying}
L. Fry, T. Hunter, M. Phanikumar, V. Fortin, and A. Gronewold,
  \emph{Identifying streamgage networks for maximizing the effectiveness of
  regional water balance modeling}, Water Resources Research 49 (2013), pp.
  2689--2700.

\bibitem{gelmanIDBayes}
A. Gelman, \emph{How to think about {“identifiability”} in {B}ayesian
  inference},
  \url{http://andrewgelman.com/2014/02/12/think-identifiability-bayesian-inference/}
  (2014).

\bibitem{gelman1992inference}
A. Gelman and D.B. Rubin, \emph{Inference from iterative simulation using
  multiple sequences}, Statistical Science 7 (1992), pp. 457--472.

\bibitem{gelman2013philosophy}
A. Gelman and C.R. Shalizi, \emph{Philosophy and the practice of {B}ayesian
  statistics}, British Journal of Mathematical and Statistical Psychology 66
  (2013), pp. 8--38.

\bibitem{gibson2006hydroclimatic}
J. Gibson, T. Prowse, and D. Peters, \emph{Hydroclimatic controls on water
  balance and water level variability in {G}reat {S}lave {L}ake}, Hydrological
  Processes 20 (2006), pp. 4155--4172.

\bibitem{gill2007partial}
J. Gill, \emph{Is partial-dimension convergence a problem for inferences from
  {MCMC} algorithms?}, Political Analysis 16 (2007), pp. 153--178.

\bibitem{gronewold2016HydroDrivers}
A.D. Gronewold, J. Bruxer, D. Durnford, J.P. Smith, A.H. Clites, F. Seglenieks,
  S.S. Qian, T.S. Hunter, and V. Fortin, \emph{Hydrological drivers of
  record-setting water level rise on {E}arth's largest lake system}, Water
  Resources Research 52 (2016), pp. 4026--4042.

\bibitem{GLWLSurgeEOS}
A.D. Gronewold, A.H. Clites, J. Bruxer, K. Kompoltowicz, J.P. Smith, T.S.
  Hunter, and C. Wong, \emph{{Water levels surge on Great Lakes}}, Eos,
  Transactions American Geophysical Union 96 (2015), pp. 14--17.

\bibitem{dashPaperOne}
A.D. Gronewold, A.H. Clites, J.P. Smith, and T.S. Hunter, \emph{A dynamic
  graphical interface for visualizing projected, measured, and reconstructed
  surface water elevations on the {E}arth's largest lakes}, Environmental
  Modelling and Software 49 (2013), pp. 34--39.

\bibitem{gron:fort:etal:2013}
A.D. Gronewold, V. Fortin, B.M. Lofgren, A.H. Clites, C.A. Stow, and F.H.
  Quinn, \emph{{Coasts, water levels, and climate change: A Great Lakes
  perspective}}, Climatic Change 120 (2013), pp. 697--711.

\bibitem{guo2002macro}
S. Guo, J. Wang, L. Xiong, A. Ying, and D. Li, \emph{A macro-scale and
  semi-distributed monthly water balance model to predict climate change
  impacts in {C}hina}, Journal of Hydrology 268 (2002), pp. 1--15.

\bibitem{holman2012improving}
K. Holman, A. Gronewold, M. Notaro, and A. Zarrin, \emph{Improving historical
  precipitation estimates over the lake superior basin}, Geophysical Research
  Letters 39 (2012).

\bibitem{hunter2015development}
T.S. Hunter, A.H. Clites, K.B. Campbell, and A.D. Gronewold, \emph{Development
  and application of a {N}orth {A}merican {G}reat {L}akes hydrometeorological
  database — {P}art {I}: {P}recipitation, evaporation, runoff, and air
  temperature}, Journal of Great Lakes Research 41 (2015), pp. 65--77.

\bibitem{husak2007use}
G.J. Husak, J. Michaelsen, and C. Funk, \emph{Use of the gamma distribution to
  represent monthly rainfall in {A}frica for drought monitoring applications},
  International Journal of Climatology 27 (2007), pp. 935--944.

\bibitem{thermoModel}
T.E.C. II and R.A. Assel, \emph{A one-dimensional ice thermodynamics model for
  the {L}aurentian {G}reat {L}akes}, Water Resources Research 30 (1994), pp.
  625--639.

\bibitem{jin2000application}
K.R. Jin, J.H. Hamrick, and T. Tisdale, \emph{Application of three-dimensional
  hydrodynamic model for {L}ake {O}keechobee}, Journal of Hydraulic Engineering
  126 (2000), pp. 758--771.

\bibitem{jin2010parameter}
X. Jin, C.Y. Xu, Q. Zhang, and V. Singh, \emph{Parameter and modeling
  uncertainty simulated by glue and a formal bayesian method for a conceptual
  hydrological model}, Journal of Hydrology 383 (2010), pp. 147--155.

\bibitem{joseph2013using}
J. Joseph and J.H. Guillaume, \emph{Using a parallelized {MCMC} algorithm in
  {R} to identify appropriate likelihood functions for {SWAT}}, Environmental
  Modelling \& Software 46 (2013), pp. 292--298.

\bibitem{kebede2006water}
S. Kebede, Y. Travi, T. Alemayehu, and V. Marc, \emph{Water balance of {L}ake
  {T}ana and its sensitivity to fluctuations in rainfall, {B}lue {N}ile basin,
  {E}thiopia}, Journal of Hydrology 316 (2006), pp. 233--247.

\bibitem{kim1996influence}
C. Kim and J. Stricker, \emph{Influence of spatially variable soil hydraulic
  properties and rainfall intensity on the water budget}, Water Resources
  Research 32 (1996), pp. 1699--1712.

\bibitem{kruschke2013posterior}
J.K. Kruschke, \emph{Posterior predictive checks can and should be {B}ayesian:
  {C}omment on {G}elman and {S}halizi, `{P}hilosophy and the practice of
  {B}ayesian statistics'}, British Journal of Mathematical and Statistical
  Psychology 66 (2013), pp. 45--56.

\bibitem{lespinas2015performance}
F. Lespinas, V. Fortin, G. Roy, P. Rasmussen, and T. Stadnyk, \emph{Performance
  evaluation of the {C}anadian {P}recipitation {A}nalysis {(CaPA)}}, Journal of
  Hydrometeorology 16 (2015), pp. 2045--2064.

\bibitem{li2007lake}
X.Y. Li, H.Y. Xu, Y.L. Sun, D.S. Zhang, and Z.P. Yang, \emph{Lake-level change
  and water balance analysis at {L}ake {Q}inghai, {W}est {C}hina during recent
  decades}, Water Resources Management 21 (2007), pp. 1505--1516.

\bibitem{lunn2009bugs}
D. Lunn, D. Spiegelhalter, A. Thomas, and N. Best, \emph{The {BUGS} project:
  Evolution, critique and future directions}, Statistics in {M}edicine 28
  (2009), pp. 3049--3067.

\bibitem{lunn2000winbugs}
D.J. Lunn, A. Thomas, N. Best, and D. Spiegelhalter, \emph{Win{BUGS}-a
  {B}ayesian modelling framework: concepts, structure, and extensibility},
  Statistics and Computing 10 (2000), pp. 325--337.

\bibitem{makhlouf1994two}
Z. Makhlouf and C. Michel, \emph{A two-parameter monthly water balance model
  for {F}rench watersheds}, Journal of Hydrology 162 (1994), pp. 299--318.

\bibitem{malve2007bayesian}
O. Malve, M. Laine, H. Haario, T. Kirkkala, and J. Sarvala, \emph{Bayesian
  modelling of algal mass occurrences—using adaptive {MCMC} methods with a
  lake water quality model}, Environmental Modelling \& Software 22 (2007), pp.
  966--977.

\bibitem{urbanizationNewFrontier}
G. Martine, G. McGranahan, M. Montgomery, and R. Fernandez-Castilla, \emph{The
  New Global Frontier: Urbanization, Poverty, and Environment in the 21st
  Century}, Routledge Earthscan, New York, NY, 2008.

\bibitem{mill:beta:falk:etal:2008}
P.C. Milly, J. Betancourt, M. Falkenmark, R.M. Hirsch, Z.W. Kundzewicz, D.P.
  Lettenmaier, and R.J. Stouffer, \emph{Stationarity is dead: whither water
  management?}, Science 319 (2008), pp. 573--574.

\bibitem{montgomery2008design}
D.C. Montgomery, \emph{Design and {A}nalysis of {E}xperiments}, John Wiley \&
  Sons, 2008.

\bibitem{mouelhi2006stepwise}
S. Mouelhi, C. Michel, C. Perrin, and V. Andr{\'e}assian, \emph{Stepwise
  development of a two-parameter monthly water balance model}, Journal of
  Hydrology 318 (2006), pp. 200--214.

\bibitem{muel:abad:garc:etal:2007}
D.S. Mueller, J.D. Abad, C.M. Garcia, J.W. Gartner, M.H. Garcia, and K.A.
  Oberg, \emph{{Errors in acoustic Doppler profiler velocity measurements
  caused by flow disturbance}}, Journal of Hydraulic Engineering 133 (2007),
  pp. 1411--1420.

\bibitem{constrainedKalman_pan_wood_2006}
M. Pan and E.F. Wood, \emph{Data assimilation for estimating the terrestrial
  water budget using a constrained ensemble {K}alman filter}, Journal of
  Hydrometeorology 7 (2006), pp. 534--547.

\bibitem{penman1948natural}
H.L. Penman, \emph{Natural evaporation from open water, bare soil and grass},
  in \emph{Proc. R. Soc. Lond. A}, Vol. 193. The Royal Society, 1948, pp.
  120--145.

\bibitem{piper1986water}
B. Piper, D. Plinston, and J. Sutcliffe, \emph{The water balance of {L}ake
  {V}ictoria}, Hydrological Sciences Journal 31 (1986), pp. 25--37.

\bibitem{plummer2003jags}
M. Plummer, \emph{et~al.}, \emph{{JAGS}: A program for analysis of {B}ayesian
  graphical models using {G}ibbs sampling}, in \emph{Proceedings of the 3rd
  international workshop on distributed statistical computing}, Vol. 124.
  Vienna, 2003, p. 125.

\bibitem{quin:guer:1986}
F.H. Quinn and B. Guerra, \emph{{Current perspectives on the Lake Erie water
  balance}}, Journal of Great Lakes Research 12 (1986), pp. 109--116.

\bibitem{team2013r}
 {R Core Team}, \emph{R: A language and environment for statistical computing},
  R Foundation for Statistical Computing, Vienna, Austria (2014).

\bibitem{raes2006simulation}
D. Raes, S. Geerts, E. Kipkorir, J. Wellens, and A. Sahli, \emph{Simulation of
  yield decline as a result of water stress with a robust soil water balance
  model}, Agricultural Water Management 81 (2006), pp. 335--357.

\bibitem{rannala2002identifiability}
B. Rannala, \emph{Identifiability of parameters in mcmc bayesian inference of
  phylogeny}, Systematic Biology 51 (2002), pp. 754--760.

\bibitem{renard2010understanding}
B. Renard, D. Kavetski, G. Kuczera, M. Thyer, and S.W. Franks,
  \emph{Understanding predictive uncertainty in hydrologic modeling: The
  challenge of identifying input and structural errors}, Water Resources
  Research 46 (2010).

\bibitem{rodell2004basin}
M. Rodell, J. Famiglietti, J. Chen, S. Seneviratne, P. Viterbo, S. Holl, and C.
  Wilson, \emph{Basin scale estimates of evapotranspiration using {GRACE} and
  other observations}, Geophysical Research Letters 31 (2004).

\bibitem{sheffield2009closing}
J. Sheffield, C.R. Ferguson, T.J. Troy, E.F. Wood, and M.F. McCabe,
  \emph{Closing the terrestrial water budget from satellite remote sensing},
  Geophysical Research Letters 36 (2009).

\bibitem{2017arXiv171010161S}
J.P. {Smith} and A.D. {Gronewold}, \emph{{Development and analysis of a
  Bayesian water balance model for large lake systems}}, ArXiv e-prints
  (2017).

\bibitem{dashPaperThree}
J.P. Smith, T.S. Hunter, A.H. Clites, C.A. Stow, T. Slawecki, G.C. Muhr, and
  A.D. Gronewold, \emph{An expandable web-based platform for visually analyzing
  basin-scale hydro-climate time series data}, Environmental Modelling \&
  Software 78 (2016), pp. 97--105.

\bibitem{spiegelhalter2014deviance}
D.J. Spiegelhalter, N.G. Best, B.P. Carlin, and A. Linde, \emph{The deviance
  information criterion: 12 years on}, Journal of the Royal Statistical
  Society: Series B (Statistical Methodology) 76 (2014), pp. 485--493.

\bibitem{spiegelhalter2002bayesian}
D.J. Spiegelhalter, N.G. Best, B.P. Carlin, and A. Van Der~Linde,
  \emph{Bayesian measures of model complexity and fit}, Journal of the Royal
  Statistical Society: Series B (Statistical Methodology) 64 (2002), pp.
  583--639.

\bibitem{swenson2009monitoring}
S. Swenson and J. Wahr, \emph{Monitoring the water balance of {L}ake
  {V}ictoria, {E}ast {A}frica, from space}, Journal of Hydrology 370 (2009),
  pp. 163--176.

\bibitem{thom:1958}
H.C. Thom, \emph{{A note on the gamma distribution}}, Monthly Weather Review 86
  (1958), pp. 117--122.

\bibitem{usgsGauges}
 {United States Geological Survey}, \emph{National {W}ater {I}nformation
  {S}ervice}, Available at https://waterdata.usgs.gov/nwis/inventory
  (2017/02/20) (2017).

\bibitem{modellingWithStakeHolders}
A. Voinov and F. Bousquet, \emph{Modelling with stakeholders}, Environmental
  Modelling and Software 25 (2010), pp. 1268--1281.

\bibitem{vorosmarty1998potential}
C.J. V{\"o}r{\"o}smarty, C.A. Federer, and A.L. Schloss, \emph{Potential
  evaporation functions compared on {US} watersheds: Possible implications for
  global-scale water balance and terrestrial ecosystem modeling}, Journal of
  Hydrology 207 (1998), pp. 147--169.

\bibitem{vorosmarty2000global}
C.J. V{\"o}r{\"o}smarty, P. Green, J. Salisbury, and R.B. Lammers, \emph{Global
  water resources: vulnerability from climate change and population growth},
  science 289 (2000), pp. 284--288.

\bibitem{xu1998reviewOfWaterBalanceMods}
C.Y. Xu and V.P. Singh, \emph{A review on monthly water balance models for
  water resources investigations}, Water Resources Management 12 (1998), pp.
  20--50.

\end{thebibliography}

\end{document}